\documentclass[structabstract,a4paper]{aa}
\usepackage{graphicx,txfonts,latexsym}
\usepackage{amssymb}
\usepackage{subfigure}
\usepackage[latin1]{inputenc}
\usepackage{natbib}
\bibpunct{(}{)}{;}{a}{}{,}

\begin{document}

\title{Analytic solutions for Navarro--Frenk--White lens models in the strong lensing regime for low characteristic convergences}
\author{H.~S.~D\'umet-Montoya\thanks{e-mail address: hdumetm@cbpf.br}
\and G.~B.~Caminha  \and M. ~Makler}
\institute{Instituto de Cosmologia, Relatividade e Astrof\'isica --- ICRA,
Centro Brasileiro de Pesquisas F\'isicas, 
Rua Dr. Xavier Sigaud 150, CEP 22290-180, Rio de Janeiro, RJ, Brazil\\
}

\abstract
{The Navarro--Frenk--White (NFW) density profile is often used to model gravitational lenses. 
For $\kappa_s \lesssim 0.1$ (where $\kappa_s$ is a parameter that defines the normalization of the NFW lens potential) --- corresponding to galaxy and galaxy group mass scales --- high numerical precision is required to accurately compute several quantities in the strong lensing regime.}
{We obtain analytic solutions for several lensing quantities for circular NFW models and their elliptical (ENFW) and pseudo-elliptical (PNFW) extensions, on the typical scales where gravitational arcs are expected to be formed, in the $\kappa_s \lesssim 0.1$ limit, by establishing their domain of validity. }
{We approximate the deflection angle of the circular NFW model and derive analytic expressions for the convergence and shear for the PNFW and ENFW models. We obtain the constant distortion curves (including the tangential critical curve), which are used to define the domain of validity of the approximations, by employing a figure-of-merit to compare with the the exact numerical solutions.
We compute the deformation cross section as a further check of the validity of the approximations.}
{We derive analytic solutions for iso-convergence contours and constant distortion curves for the models considered here. We also obtain the deformation cross section, which is given in closed form for the circular NFW model and in terms of a one-dimensional integral for the elliptical ones. In addition, we provide a simple expression for the ellipticity of the iso-convergence contours of the pseudo-elliptical models and the connection of characteristic convergences among the PNFW and ENFW models.}
{We conclude that the set of solutions derived here is generally accurate for $\kappa_s \lesssim 0.1$. For 
low ellipticities, values up to $\kappa_s \simeq 0.18$ are allowed. On the other hand, the mapping among PNFW and the ENFW models is valid up to $\kappa_s \simeq 0.4$.
The solutions derived in this work can be used to speed up numerical codes and ensure their accuracy in the low 
$\kappa_s$ regime, including applications to arc statistics and other strong lensing observables.
}

\titlerunning{Analytic Solutions for NFW lens models in the Strong Lensing Regime}
\authorrunning{D\'umet-Montoya et al.}
\maketitle

\section{Introduction: \label{introd}}

Gravitational arcs are powerful tools to probe the mass distribution in galaxies  \citep{2009ApJ...703L..51K,2011MNRAS.415.2215B,2012ApJ...750...10S} and clusters of galaxies  \citep{1989ApJ...337..621K,1993ApJ...403..497M,1997A&A...319..764H,2006ApJ...642...39C}. In addition, their abundance can help to constrain cosmological models \citep{1998A&A...330....1B,2002A&A...387..788G,2003A&A...409..449B,2005IAUS..225..185M,2010Sci...329..924J}. 

Two techniques have been used to extract information from gravitational arcs. The first is \emph{arc-statistics}: counting arcs as a function of their properties, such as the length-to-width ratio or angular separation, for a lens sample \citep{1993MNRAS.262..187W,1994ApJ...431...74G,1994A&A...287....1B, 1995A&A...297....1B, 1995A&A...299...11B}. The second is \emph{inverse modeling}: using  arcs in individual clusters or galaxies aiming to determine the mass distribution of the lens and source properties \citep{1993A&A...273..367K,2001astro.ph..2340K,2002A&A...387..788G,2006MNRAS.372.1187W,2007NJPh....9..447J,2010Sci...329..924J}.  

These approaches have motivated arc searches in wide field surveys \citep[][Erben et al., in prep]{2003ApJ...593...48G,2007ApJ...660.1176E,2007A&A...461..813C,2009MNRAS.392..104B,2010ApJ...724L.137K,kneib2010,2011AJ....141...94G,2011RAA....11.1185W,2012ApJ...749...38M,2012ApJ...744..156B,2012ApJ...761....1W}, as well as in images targeting clusters \citep{1999A&AS..136..117L,2001ApJ...553..668E, 2003ApJ...584..691Z,2005MNRAS.359..417S,2005ApJ...627...32S,2008AJ....135..664H,2010A&A...513A...8K,2011MNRAS.418...54H,2012arXiv1210.4136F,2012ApJS..199...25P} and galaxies \citep{1999AJ....117.2010R,2004ApJ...600L.155F,2006ApJ...638..703B,2006MNRAS.369.1521W,2007ApJ...660L..31M,2008MNRAS.385..918K,2008ApJS..176...19F,2008MNRAS.389.1311J}. Moreover, the upcoming wide-field surveys, such as the  Dark Energy Survey\footnote{\texttt{http://www.darkenergysurvey.org}} \citep[][Frieman et al., in prep;  Lahav et al., in prep]{2005astro.ph.10195A}, which started taking data in 2012, are expected to detect strong lensing systems in the thousands, about an order of magnitude more than the current homogeneous samples.

A widely used model for representing the radial distribution of dark matter from galaxy to cluster of galaxies mass scales is the Navarro--Frenk--White profile \citep[][hereafter NFW]{1996ApJ...462..563N, 1997ApJ...490..493N}, whose mass density is given by
\begin{equation}
\rho(r)=\frac{\rho_s}{(r/r_s)(1+r/r_s)^2}, \label{nfw_den_profile}
\end{equation}
where $r_s$ and $\rho_s$ are the scale radius and characteristic density, respectively. It is useful to define the characteristic convergence
\begin{equation}
\kappa_s = \frac{r_s \,\rho_s}{\Sigma_{\rm crit}} \label{kappa_s_nnfw}
\end{equation}
as a mass parameter, where, $\Sigma_{\rm crit}$ is the critical surface mass density \citep{SEF,2001stgl.book.....P,2002glml.book.....M}.

Some observational properties of many  arcs systems (such as arc multiplicity, relative positions, morphology) imply that the mass distribution of the lens is not axially symmetric. Furthermore, results from N-body simulations  predict that dark matter halos are typically triaxial in shape and can be modeled by ellipsoids \citep{2002ApJ...574..538J,2007MNRAS.378...55M}. A first approximation to model realistic lenses is to consider elliptical lens models, where the ellipticity is introduced either on the mass distribution \citep[the so-called elliptical models,][]{1990A&A...231...19S,1998ApJ...502..531B,2001astro.ph..2340K,2003ApJ...599....7O} or on the lensing potential \citep[the so-called pseudo-elliptical models,][]{1987ApJ...321..658B,1993ApJ...417..450K,doi:10.1142/9789812778017_0008,2002A&A...390..821G}. While elliptical models are generally more realistic, they are usually much more time-consuming for lensing calculations than are pseudo-elliptical ones. As a consequence, both have been used in the literature, depending on the application.

In this work we consider both elliptical and pseudo-elliptical lens models in which the radial mass distribution is given by the projected NFW profile. When used to represent lenses on galactic mass scales \citep[see, e.g.,][]{2000PASJ...52...99A, 2003MNRAS.344.1029D, 2009MNRAS.392..945V,2012ApJ...750...10S, 2013arXiv1302.0288L}, the characteristic convergence of the NFW model (Eq. (\ref{kappa_s_nnfw})) takes very low values. For instance, lensing systems with $M_{200} < 10^{13}M_{\sun}h^{-1}$, $z_{\rm L} \leq 1$ and sources with $z_{\rm S} = 2z_{\rm L}$, have values of \footnote{To obtain this value of $\kappa_s$ we use the expressions in \citet{CaminhaMagBias} with the same choices for the NFW and cosmological parameters as described in Sect. 2.2 of \citet{2012A&A...544A..83D}.} $\kappa_s < 0.05$. In this regime high numerical precision is required to accurately compute the functions involved in gravitational lensing, such as the deflection angle and its derivatives. Therefore, the numerical codes become either slower or, worse, they provide unreliable results as we go to low values of $\kappa_s$.

This issue has apparently not been addressed in the literature so far. A solution is to obtain analytical expressions, which are valid in this regime, providing at the same time a fast and reliable way to compute the relevant lensing functions. 
In this work, we present approximations for the deflection angle, convergence, and shear of the circular, pseudo-elliptical, and elliptical NFW models in the strong lensing regime. We use them to derive analytical solutions for iso-convergence contours, critical curves, and constant distortion curves. We compare these solutions with the exact calculations to determine a domain of validity for these approximations. Moreover, for applications to arc statistics, we apply these solutions to the calculation of the deformation cross section.

The outline of this work is as follows. In Sect. \ref{basis} we present a few basic definitions of lensing quantities and introduce the notation used in this paper. In Sect. \ref{approx_nfws} we show the approximations for the lensing functions of the circular NFW model and their extension to the pseudo-elliptical NFW (PNFW) and elliptical NFW (ENFW) models. We also derive analytical expressions for iso-convergence contours and critical curves for these models.  In Sect. \ref{dcs_sects} we obtain analytical solutions for constant distortion curves and for the deformation cross section. In Sect. \ref{val_sols} we determine a domain of validity of these solutions in terms of the characteristic convergences and ellipticity parameters. In Sect. \ref{mapping} we obtain a mapping among the PNFW and ENFW model parameters. In Sect.~\ref{s&c} we present the summary and concluding remarks. In Appendix~\ref{app_enfw} we present the expressions for the potential derivatives of elliptical models. In Appendix  \ref{app_fit} we provide fitting functions giving upper limits of  $\kappa_s$ for the validity of the analytical solutions as a function of ellipticity. 

\section{Definitions and notation\label{basis}}

We present below some basic definitions of gravitational lensing in order to set up the notation throughout this work.
More details on the subject can be found, say, in \citet{SEF}, \citet{2001stgl.book.....P}, and \citet{2002glml.book.....M} 

The lensing properties are encoded in the lens equation, which relates the observed image position $\vec{\xi}$ to a given source position $\vec{\eta}$ (both with respect to the optical axis). By defining the length scales $\xi_0$ on the lens plane and $\eta_0=\xi_0 D_{\rm OS}/D_{\rm OL}$ on the source plane, where $D_{\rm OS}$ and $D_{\rm OL}$ are the angular-diameter distances to the lens and source planes, respectively, the lens equation is given in its dimensionless form by
\begin{equation}
\vec{y}=\vec{x}-\vec{\alpha}(\vec{x}),
\label{lens_eq_adim}
\end{equation}
where $\vec{y}=\vec{\eta}/\eta_0$, $\vec{x}=\vec{\xi}/\xi_0$ and $\vec{\alpha}(\vec{x})=\left(\frac{D_{\rm OL} D_{\rm LS}}{\xi_0 D_{\rm OS}} \right)\vec{\hat{\alpha}}(\xi_0\vec{x})$, where $\hat{\alpha}$  is the 
deflection angle due to the lensing mass distribution. 

Properties of the local mapping are described by the Jacobian matrix of Eq. (\ref{lens_eq_adim})
\begin{equation}
\tens{J}_{ij}(\vec{x})=\delta_{ij}-\partial_i\alpha_j(\vec{x}).
\end{equation}
The two eigenvalues of this matrix are written as 
\begin{equation}
\lambda_r(\vec{x})=1-\kappa(\vec{x}) + \gamma(\vec{x}) \quad  \mathrm{and} \quad \lambda_t(\vec{x})=1-\kappa(\vec{x}) - \gamma(\vec{x}), \label{eigen_r-t}
\end{equation}
where 
\begin{equation}
\kappa(\vec{x}) = \frac{1}{2}\left( \partial_1\alpha_1(\vec{x})+\partial_2\alpha_2(\vec{x}) \right) \label{conv}
\end{equation}
is the convergence and $\gamma(\vec{x})=\sqrt{\gamma^2_1(\vec{x})+\gamma^2_2(\vec{x})}$, is the shear, which has components
\begin{equation}
\gamma_1(\vec{x}) =  \frac{1}{2}\left(\partial_1\alpha_1(\vec{x})-\partial_2\alpha_2(\vec{x})\right), \ \ \gamma_2 =  \frac{1}{2}\left(\partial_1\alpha_2(\vec{x})+\partial_2\alpha_1(\vec{x})\right). \label{shear1_2}
\end{equation}
Magnification in the radial  direction is given by $\lambda^{-1}_r(\vec{x})$ and in the tangential by $\lambda^{-1}_t(\vec{x})$. Points satisfying the conditions $\lambda_r(\vec{x})=0$ and $\lambda_t(\vec{x})=0$ are the radial and tangential critical curves, respectively. Mapping these curves onto the source plane, we obtain the corresponding caustics.

\section{Solutions for iso-convergence contours and critical curves\label{approx_nfws}}

In this section we present the approximations for the lensing functions of the the circular NFW (Sect. \ref{axial_nfw}), the PNFW (Sect. \ref{pe_nfw}), and the ENFW (Sect. \ref{e_nfw}) models.

\subsection{Circular NFW model \label{axial_nfw}}

Following \citet{1996A&A...313..697B}, from the density profile (\ref{nfw_den_profile}) and taking $\xi_0=r_s$, the dimensionless deflection angle is given by
\begin{equation}
\alpha(x)=\frac{4\kappa_s}{x}\left(\ln{\frac{x}{2}}+\frac{2}{\sqrt{1-x^2}}\arctan{\rm\!h}{\sqrt{\frac{1-x}{1+x}}} \right), \label{angle_nfw_tot}
\end{equation}
from which the convergence and shear are derived. In the limit of $\kappa_s \lesssim 0.1$, the typical scales corresponding to the strong lensing regime (e.g., the size of critical curve and caustics) are much smaller than  unity.
In this case high numerical precision is required to accurately compute lensing quantities, such as the shear and convergence.  However, simple analytic expressions can be found in this regime by avoiding such numerical difficulties.
Keeping the first terms in a series expansion of (\ref{angle_nfw_tot}) for $x \ll 1$ leads to
\begin{equation}
\alpha(x)=-x\kappa_s\left(1+2\ln{\frac{x}{2}}\right). \label{alpha_nfw}\\
\end{equation}
In this case, the convergence and shear are given by
\begin{eqnarray}
\kappa(x) &=& \frac{1}{2}\left(\frac{\alpha}{x}+\frac{d\alpha}{dx}\right)= -2\kappa_s\left(1+\ln{\frac{x}{2}}\right), \label{kappa_nfw}\\
\gamma(x)&=& \frac{1}{2}\left(\frac{\alpha}{x}-\frac{d\alpha}{dx}\right)=  \kappa_s. \label{shear_nfw}
\end{eqnarray}
Allowing a relative deviation of less than 1\% (0.1\%) with respect to the exact expression, we found that Eq. (\ref{alpha_nfw}) is a good approximation to the deflection angle for $x\leq 0.12$ ($x\leq 0.04$), while the same holds for Eqs. (\ref{kappa_nfw}) and (\ref{shear_nfw}) for the convergence and shear for $x\leq 0.08$ and $x\leq 0.05$  ($x\leq 0.025$ and $x\leq 0.015$), respectively.

From Eq. (\ref{kappa_nfw}) and fixing a value for the iso-convergence contour as $\kappa_{\rm const}$, the equation $\kappa(x)= \kappa_{\rm const}$ has the solution
\begin{equation}
x_\kappa = 2\exp{\left(- \frac{\kappa_{\rm const}+2\kappa_s}{ 2\kappa_s} \right)}. \label{xk_nfw}
\end{equation}

From Eqs. (\ref{eigen_r-t}), (\ref{kappa_nfw}), and (\ref{shear_nfw}) it follows that
\begin{equation}
\lambda_t(x)=1+\kappa_s\left(1+2\ln{\frac{x}{2}} \right), \quad \lambda_r(x)=1+\kappa_s\left(3+2\ln{\frac{x}{2}} \right) \label{lt-lr_nfw} 
\end{equation}
such that the determinant of the Jacobian matrix is
\begin{equation}
{\rm det}\,\tens{J}(x)=\kappa^2(x)-2\kappa(x)+1-\kappa^2_s. \label{det_nfw}
\end{equation}
The solutions for the tangential and radial critical curves follow from the equations above. The radial coordinates of these curves are\footnote{These solutions are shown in \citet{2003MNRAS.340..105M}. However, there is a typo in this reference as they appear in their Eq. (11) as the eigenvalues of the Jacobian matrix.}
\begin{eqnarray}
x_t &=& 2 \exp{\left( -\frac{1+\kappa_s}{2\kappa_s} \right)}, \label{xt_nfw}  \\
x_r &=& 2 \exp{\left( -\frac{1+3\kappa_s}{2\kappa_s} \right)}. \label{xr_nfw}
\end{eqnarray}
The expressions for the caustics are obtained straightforwardly  by inserting the expressions above in the lens equation with the deflection angle given in Eq. (\ref{alpha_nfw}). The validity of these solutions is discussed in Sect \ref{val_sols}.

\subsection{Pseudo-Elliptical NFW model \label{pe_nfw}}

The construction of pseudo-elliptical models is made by replacing the radial coordinate of the lensing potential by
\begin{equation}
x_\varphi=x\sqrt{a_{\varphi}\cos^2{\phi}+b_{\varphi} \sin^2{\phi}}\label{rad_coord_pnfw},
\end{equation}
where $a_{\varphi} $ and $b_{\varphi}$ are two parameters that define the lensing potential ellipticity. {Adopting the approach of the {\it angle deflection method} introduced by \citet{2002A&A...390..821G} \citep[see also][hereafter DCM]{2012A&A...544A..83D}, from Eqs. (\ref{alpha_nfw})--(\ref{shear_nfw}), the deflection angle, convergence, and components of the shear of the PNFW model are
\begin{eqnarray}
\vec{\alpha}_\varphi(\vec{x})&=&\alpha(x_\varphi)\left(\sqrt{a_{\varphi} }\cos{\phi_\varphi},\sqrt{b_{\varphi}}\sin{\phi_\varphi} \label{alpha_pnfw}\right), \label{aphi}\\
\kappa_\varphi(\vec{x}) &=&\mathcal{A}\kappa(x_\varphi)-\mathcal{B}\kappa_s^\varphi\cos{2\phi_\varphi}, \label{kappa_pnfw}\\
\gamma_{1\varphi}(\vec{x})&=& \mathcal{B}\kappa(x_\varphi)-\mathcal{A}\kappa_s^\varphi\cos{2\phi_\varphi}, \label{g1_pnfw}\\
\gamma_{2\varphi}(\vec{x})&=& -\sqrt{a_{\varphi}  b_{\varphi} }\kappa_s^\varphi\sin{2\phi_\varphi}, \label{g2_pnfw}
\end{eqnarray}
where 
\begin{equation}
\mathcal{A}=\frac{1}{2}(a_{\varphi} +b_{\varphi} ),\ \mathcal{B}=\frac{1}{2}(a_{\varphi} -b_{\varphi} ) \ \textrm{and} \ \phi_\varphi=\arctan{\left(\sqrt{b_{\varphi} /a_{\varphi} }\tan{\phi}\right)},
\label{phivarphi}
\end{equation}
and we denote the NFW characteristic convergence, Eq. (\ref{kappa_s_nnfw}), by $\kappa_s^\varphi$.

From Eq. (\ref{kappa_pnfw}) and fixing a value for the iso-convergence contour as $\kappa_{\rm const}$, the equation $\kappa_\varphi(\vec{x})= \kappa_{\rm const}$ has the solution
\begin{equation}
x_{\kappa}(\phi)=\frac{2}{\sqrt{\mathcal{A}+\mathcal{B}\cos{2\phi}}}\exp{\left(-\frac{\kappa_{\rm const}+(2\mathcal{A}+\mathcal{B}\cos{2\phi_\varphi})\kappa^\varphi_s}{2\mathcal{A}\kappa^\varphi_s} \right)}.\label{xk_pnfw}
\end{equation}
Therefore, the iso-convergence contours are not elliptical, as is well known for pseudo-elliptical models.

The solutions for critical curves are a bit more involved. From Eqs. (\ref{kappa_pnfw})--(\ref{g2_pnfw}), the determinant of the Jacobian matrix is
\begin{equation}
{\rm det}\,\tens{J}(\vec{x})=a_{\varphi}  b_{\varphi}  \kappa^2(x_\varphi)-2\mathcal{A}\kappa(x_\varphi)+1+2\mathcal{B}\kappa_s^\varphi\cos{2\phi_\varphi}-a_{\varphi} b_{\varphi} (\kappa_s^\varphi)^2.
\label{jacob_pnfw}
\end{equation}
Then, solving the equation ${\rm det}\,\tens{J}(\vec{x})=0$ for $\kappa(x_\varphi)$, defining
\begin{displaymath}
\tilde{\kappa}^{\pm}(\phi)= (a_{\varphi}  b_{\varphi} )^{-1}\left(\mathcal{A}\pm \sqrt{\mathcal{B}^2+(a_{\varphi} b_{\varphi} \kappa_s^\varphi)^2-2a_{\varphi} b_{\varphi} \mathcal{B}\kappa_s^\varphi\cos{2\phi_\varphi}} \right),
\end{displaymath}
and inverting $\kappa(x_\varphi)= \tilde{\kappa}^{\pm}(\phi)$, using Eq. (\ref{rad_coord_pnfw}), we obtain
\begin{eqnarray}
x_{t}(\phi)&=&\frac{2}{\sqrt{\mathcal{A}+\mathcal{B}\cos{2\phi}}}\exp{\left(-\frac{\tilde{\kappa}^{-}(\phi)+2\kappa^\varphi_s}{2\kappa^\varphi_s}  \right)},\label{xt_pnfw} \\ 
x_{r}(\phi) &=& \frac{2}{\sqrt{\mathcal{A}+\mathcal{B}\cos{2\phi}}}\exp{\left(-\frac{\tilde{\kappa}^{+}(\phi)+2\kappa^\varphi_s}{2\kappa^\varphi_s}  \right)}.\label{xr_pnfw}
\end{eqnarray}
These curves are mapped onto the source plane by using the lens equation with the deflection angle given in Eq. (\ref{alpha_pnfw}). The validity of these solutions is discussed in Sect \ref{val_sols}.

\subsection{Elliptical NFW model \label{e_nfw}}

We construct the ENFW model by replacing the radial coordinate of the surface mass density, Eq. (\ref{kappa_nfw}), by 
\begin{equation}
x_\Sigma=x\sqrt{\cos^2{\phi}/a^2_\Sigma + \sin^2{\phi}/b^2_\Sigma},\label{coord_enfw}
\end{equation}
where $a_{\Sigma}$ and $b_{\Sigma}$ define the ellipticity of the mass distribution.

The lensing functions of this model can be written as (see Appendix \ref{app_enfw})
\begin{eqnarray}
\alpha_{1\Sigma}(\vec{x}) & = & a_{\Sigma}b_{\Sigma} x\left[
\mathcal{J}_0\kappa(x)-2\kappa^\Sigma_s\mathcal{L}_0(\phi)\right] \cos{\phi}, \label{alpha1_enfw} \\ 
\alpha_{2\Sigma}(\vec{x}) & = & a_{\Sigma}b_{\Sigma}  x \left[\mathcal{J}_1\kappa(x)-2\kappa^\Sigma_s\mathcal{L}_1(\phi)\right]
\sin{\phi}, \label{alpha2_enfw} \\ 
\kappa_\Sigma(\vec{x})& \equiv &\kappa(x_\Sigma)=1-\frac{1}{2}\left[\mathcal{P} +\mathcal{Q}-a_{\Sigma}b_{\Sigma}(\mathcal{J}_0+\mathcal{J}_1)\kappa(x) \right], \label{kappa_enfw}\\
\gamma_{1\Sigma}(\vec{x})&=&\frac{1}{2}\left[a_{\Sigma}b_{\Sigma}  (\mathcal{J}_0-\mathcal{J}_1)\kappa(x)+\mathcal{Q}-\mathcal{P}\right], \label{g1_enfw}\\
\gamma_{2\Sigma}(\vec{x})&=&-a_{\Sigma}b_{\Sigma}\kappa_s^\Sigma\sin{2\phi}\mathcal{K}_1(\phi), \label{g2_enfw}
\end{eqnarray}
where we denote the characteristic convergence, Eq. (\ref{kappa_s_nnfw}), by $\kappa_s^\Sigma$ and define  
\begin{eqnarray}
\mathcal{P}&=&1+2a_{\Sigma}b_{\Sigma}\kappa_s^\Sigma\left[\mathcal{K}_0(\phi)\cos^2{\phi}+\mathcal{L}_0(\phi) \right], \label{p_enfw}\\
\mathcal{Q}&=&1+2a_{\Sigma}b_{\Sigma}\kappa_s^\Sigma \left[\mathcal{K}_2(\phi)\sin^2{\phi}+\mathcal{L}_1(\phi) \right]. \label{q_enfw}
\end{eqnarray}
We computed the accuracy of Eqs. (\ref{alpha1_enfw}) and (\ref{alpha2_enfw}) with respect to the exact expressions (Eq. (\ref{d1-2_ellip})) for the angles $ \phi$ in which the deviations are maximal. For a percentile deviation less than 1\% (0.1\%), such expressions are good approximations of the exact components of the deflection angle for  $x\leq 0.11$ ($x\leq 0.03$) and $x\leq 0.10$ ($x\leq 0.025$), respectively, within the ellipticity parameter range $0.1$--$0.6$.

For an iso-convergence contour value $\kappa_{\rm const}$, the solution for $\kappa_\Sigma(\vec{x})=\kappa_{\rm const}$ is
\begin{equation}
x_{\kappa}(\phi)=\frac{2}{\sqrt{\cos^2{\phi}/a^2_{\Sigma}+\sin^2{\phi}/b^2_{\Sigma}}}\exp{\left(-\frac{\kappa_{\rm const}+2\kappa_s^\Sigma}{2\kappa_s^\Sigma} \right)},
\label{xk_enfw}
\end{equation}
 which are ellipses, as expected.

From the definitions in (\ref{eigen_r-t}) and Eqs. (\ref{kappa_enfw})--(\ref{g2_enfw}), the determinant of the Jacobian matrix is
\begin{equation}
{\rm det}\,\tens{J}(\vec{x})=(a_{\Sigma} b_{\Sigma})^2\mathcal{J}_0\mathcal{J}_1\kappa^2(x)-a_{\Sigma} b_{\Sigma}(\mathcal{J}_1\mathcal{P}+\mathcal{J}_0\mathcal{Q})\kappa(x)+\mathcal{P}\mathcal{Q}-\gamma^2_{2\Sigma}. \label{jacob_enfw}
\end{equation}
Following a similar procedure to obtain the critical curves for the PNFW model, the solutions of the equation  ${\rm det}\,\tens{J}(\vec{x})=0$ are
\begin{eqnarray}
x_t(\phi) &=& 2\exp{\left(-\frac{\tilde{\kappa}^{-}_\Sigma(\phi) +2\kappa_s^\Sigma}{2\kappa_s^\Sigma} \right)}, \label{xt_enfw}\\
x_r(\phi) &=& 2\exp{\left(-\frac{\tilde{\kappa}^{+}_\Sigma(\phi) +2\kappa_s^\Sigma}{2\kappa_s^\Sigma} \right)},\label{xr_enfw}
\end{eqnarray}
where we define
\begin{displaymath}
\tilde{\kappa}^{\pm}_\Sigma(\phi)= \frac{\mathcal{J}_1\mathcal{P}+\mathcal{J}_0 \mathcal{Q}\pm \sqrt{(\mathcal{J}_1\mathcal{P}-\mathcal{J}_0\mathcal{Q})^2+4\mathcal{J}_1\mathcal{J}_0\gamma^2_{2\Sigma}}}{2 a_{\Sigma} b_{\Sigma}\mathcal{J}_0\mathcal{J}_1}.
\end{displaymath}
The corresponding caustics are obtained by using the lens equation with the deflection angle given in Eqs. (\ref{alpha1_enfw}) and (\ref{alpha2_enfw}). The validity of these solutions
is discussed in Sect \ref{val_sols}.

\section{Solutions for the deformation cross section \label{dcs_sects}}

Gravitational arcs are usually defined as images with length-to-width ratio, $L/W$, greater than a threshold $R_{\rm th}$. For fast calculations in arc statistics, it is useful to approximate $L/W$ to the ratio of the eigenvalues of the Jacobian matrix \citep{1993MNRAS.262..187W,1994A&A...287....1B,1997MNRAS.286L...7H}
\begin{equation}
\frac{L}{W} \simeq \left|  R_\lambda \right |, 
\label{lw}
\end{equation}
where  $ R_\lambda = \lambda_r/\lambda_t$. This approximation holds for infinitesimal circular sources and breaks down for arcs generated by the merger of multiple images \citep{2008ApJ...687...22R} or by large or noncircular sources. 

In this section, using the approximation above, we derive analytical solutions for constant distortion curves for the NFW models (Sect. \ref{cdc_nfws}). We thereafter employ these solutions to compute the arc cross section (Sect. \ref{dcs_subsect}).

\subsection{Constant distortion curves \label{cdc_nfws}}

A typical arc-forming region in the lens plane is determined by the so-called constant distortion curves, corresponding to the $|R_\lambda|= R_{\rm th}$ contours. An often used value for $R_{\rm th}$ is $10$. As the value of $R_{\rm th}$ is decreased, the inner curve (corresponding to $R_\lambda=-R_{\rm th}$) gets closer to the center, while the outer enclosing curve ($R_\lambda=+R_{\rm th}$) reaches higher radii, where the analytic approximations derived in section 3 are less accurate. Therefore by using a lower value of $R_{\rm th}$ to determine the limit of validity of these approximations, we are assuring they are even more accurate in the arc formation region. For this reason, we adopt $R_{\rm th}=5$ when we make the numerical comparisons to the exact solution throughout this paper.

From the approximations given in Sect. \ref{approx_nfws}, it is possible to obtain analytical solutions for the radial coordinates of constant distortion curves. For the circular NFW model, from Eq. (\ref{lt-lr_nfw}), the equation $R_\lambda(\vec{x})= R_{\rm th}$ has the solution
\begin{equation}
x_\lambda=2\exp{\left(\frac{1+3\kappa_s-(1+\kappa_s)R_{\rm th}}{2\kappa_s(R_{\rm th}-1)} \right)}.
\label{cdc_nfw}
\end{equation}

For the PNFW model, calculating the radial coordinates of the constant distortion curves is a bit more complicated.
From Eqs. (\ref{kappa_pnfw})--(\ref{g2_pnfw}), solving the equation $R^2_\lambda(\vec{x})= R^2_{\rm th}$ for $\kappa(x_\varphi)$, we obtain 
\begin{equation}
\kappa(x_\varphi)=\frac{\tilde{\kappa}_{\lambda \varphi}}{2\left(\mathcal{A}^2-\mathcal{B}^2 Q^2_{\rm th}\right)},\label{prev_rl_pnfw}
\end{equation}
where
\begin{displaymath}
\tilde{\kappa}_{\lambda \varphi} = \mathcal{R}_{1\varphi} +\mathcal{R}_{2\varphi} -\sqrt{(\mathcal{R}_{1\varphi}-\mathcal{R}_{2\varphi})^2+4(\mathcal{A}^2-\mathcal{B}^2 Q^2_{\rm th})Q^2_{\rm th} \gamma^2_{2\varphi}}
\end{displaymath}
with 
\begin{eqnarray*}
Q_{\rm th} &= &\frac{R_{\rm th}+1}{R_{\rm th}-1}, \\ 
\mathcal{R}_{1\varphi} &=& (\mathcal{A}-\mathcal{B}Q_{\rm th})(1+(\mathcal{A}Q_{\rm th}+\mathcal{B})\kappa^\varphi_s\cos{2\phi_\varphi}),\\
\mathcal{R}_{2\varphi} &=& (\mathcal{A}+\mathcal{B}Q_{\rm th})(1-(\mathcal{A}Q_{\rm th}-\mathcal{B})\kappa^\varphi_s\cos{2\phi_\varphi}),
\end{eqnarray*}
where $\phi_{\varphi}$ is given in Eq. (\ref{phivarphi}). Inverting Eq. (\ref{prev_rl_pnfw}) and using Eq. (\ref{rad_coord_pnfw}), we obtain for any angular position
\begin{equation}
x_{\lambda \varphi} =\frac{2}{\sqrt{\mathcal{A}+\mathcal{B}\cos{2\phi}}}\exp{\left(-\frac{\tilde{\kappa}_{\lambda\varphi}+4(\mathcal{A}^2-\mathcal{B}^2\mathcal{Q}^2_{\rm th})\kappa^\varphi_s}{4(\mathcal{A}^2-\mathcal{B}^2\mathcal{Q}^2_{\rm th})\kappa^\varphi_s}\right)}. \label{cdc_pnfw}
\end{equation} 
 
Following the same procedure as above, for the ENFW model, from Eqs. (\ref{kappa_enfw})--(\ref{g2_enfw}), we obtain for each angular position  
\begin{equation}
x_{\lambda \Sigma}=2\exp{\left(-\frac{\tilde{\kappa}_{\lambda \Sigma} + 4\mathcal{Q}_{1\Sigma}\mathcal{Q}_{2\Sigma}\kappa^\Sigma_s}{4\mathcal{Q}_{1\Sigma}\mathcal{Q}_{2\Sigma}\kappa^\Sigma_s}\right)},
\label{cdc_enfw}
\end{equation}
where we have defined
\begin{eqnarray*}
\tilde{\kappa}_{\lambda \Sigma}&=& \mathcal{R}_{1\Sigma}\mathcal{Q}_{2\Sigma}+\mathcal{R}_{2\Sigma}\mathcal{Q}_{1\Sigma} - \\ 
& & \sqrt{(\mathcal{R}_{1\Sigma}\mathcal{Q}_{2\Sigma} -\mathcal{R}_{2\Sigma}\mathcal{Q}_{1\Sigma})^2+16 \mathcal{Q}_{1\Sigma}\mathcal{Q}_{2\Sigma}Q^2_{\rm th}\gamma^2_{2\Sigma}}
\end{eqnarray*}
with
\begin{eqnarray*}
\mathcal{R}_{1\Sigma} &=& \mathcal{P}+\mathcal{Q}+Q_{\rm th}(\mathcal{P}-\mathcal{Q}), \\ 
\mathcal{R}_{2\Sigma} &=& \mathcal{P}+\mathcal{Q}-Q_{\rm th}(\mathcal{P}-\mathcal{Q}), \\
Q_{1\Sigma} &=& a_\Sigma b_\Sigma[\mathcal{J}_0+\mathcal{J}_1+Q_{\rm th}(\mathcal{J}_0-\mathcal{J}_1)],\\
Q_{2\Sigma}& =& a_\Sigma b_\Sigma[\mathcal{J}_0+\mathcal{J}_1-Q_{\rm th}(\mathcal{J}_0-\mathcal{J}_1)],
\end{eqnarray*}
where $\mathcal{P}$, $\mathcal{Q}$, and $\mathcal{J}_n$ are given in Eqs. (\ref{p_enfw}), (\ref{q_enfw}), and (\ref{an_enfw}), respectively.

 In Eqs. (\ref{cdc_nfw}), (\ref{cdc_pnfw}), and (\ref{cdc_enfw}) the solution for the equation $R_\lambda(\vec{x})=-R_{\rm th}$ is obtained by replacing $R_{\rm th} \rightarrow -R_{\rm th}$. Also, at the limits $R_{\rm th} \rightarrow \infty$ and $R_{\rm th} \rightarrow 0$, these expressions yield the radial coordinates of the tangential (Eqs. (\ref{xt_nfw}), (\ref{xt_pnfw}), and (\ref{xt_enfw})) and radial (Eqs. (\ref{xr_nfw}), (\ref{xr_pnfw}), and (\ref{xr_enfw})) critical curves of the corresponding models.

\subsection{Arc cross section \label{dcs_subsect}}

The arc cross section, $\sigma_{R_{\rm th}}$, is defined as the weighted area in the source plane, such that sources within it will be mapped into arcs with $L/W \geq R_{\rm th}$. This cross section 
is usually computed using a large sample of arcs obtained from ray-tracing an even larger number of finite sources and is 
computationally demanding. 
An alternative for fast calculations is to use the approximation (\ref{lw}). In this case, the arc cross section is calculated in the lens plane as \citep[][DCM]{2006A&A...447..419F}
\begin{eqnarray}
\sigma_{R_{\rm th}}= \left(\frac{\xi_0}{D_{\rm OL}}\right)^2\tilde{\sigma}_{R_{\rm th}}; \ \tilde{\sigma}_{R_{\rm th}}=\int_{|R_\lambda|\geq R_{\rm th}}|{\rm det}\,\tens{J}(\vec{x})| xdxd\phi,
\label{sigma}
\end{eqnarray}
where the quantity $\tilde{\sigma}_{R_{\rm th}}$ is known as (dimensionless) deformation cross section.

For low values of the characteristic convergences, the determinant of the Jacobian (see Eqs. (\ref{det_nfw}), (\ref{jacob_pnfw}), and (\ref{jacob_enfw})) takes the form
\begin{displaymath}
{\rm det}\,\tens{J}(\vec{x}) = A \kappa^2(\vec{x}) + B \kappa(\vec{x}) + C,
\end{displaymath}
where $A$, $B$, and $C$ are independent of the radial coordinates. It is possible to reduce the calculation of (\ref{sigma}) to a one-dimensional integral. Inserting the expression above into Eq. (\ref{sigma}) and integrating over the radial coordinate, within the lower and upper limits given by the constant distortion curves (i.e., from $-R_{\rm th}$ to $R_{\rm th}$), we obtain
\begin{equation}
\tilde{\sigma}_{R_{\rm th}}=\frac{1}{2}\int_{0}^{2\pi}\left[\mathcal{S}(x_{+})+\mathcal{S}(x_{-})-2\mathcal{S}(x_t)\right]d\phi \label{sigma_medio},
\end{equation}
where $\mathcal{S}$, is a function (given below) resulting from the integration of ${\rm det}\,\tens{J}(\vec{x})$ over the radial coordinate, $x_{\pm} = x_{\pm}(\phi)$ are the solutions for the radial coordinate of the constant distortion curves for $-R_{\rm th}$ and $R_{\rm th}$,  and $x_t = x_t(\phi)$ is the solution for the radial coordinate of the tangential critical curve.

For the circular NFW model we have
\begin{equation}
\mathcal{S}=x^2\left[1-\kappa_s-\kappa(x)\right]^2,
\end{equation}
such that Eq. (\ref{sigma_medio}) gives
\begin{equation}
\tilde{\sigma}_{R_{\rm th}}=\frac{4\pi\kappa^2_s}{(R^2_{\rm th}-1)^2}\left[(R_{\rm th}+1)^2x^2_{+}+(R_{\rm th}-1)^2x^2_{-} \right].
\label{dcs_nfw_analytic}
\end{equation}
where $x_{\pm}$ are given in Eq. (\ref{cdc_nfw}) for $R_{\rm th}$ and $-R_{\rm th}$, respectively. This expression shows the exponential dependence of the cross section on $\kappa_s$  \citep{CaminhaMagBias}. For instance, varying $\kappa_s$ from $0.01$ to $0.1$, $\tilde{\sigma}_{5}$ changes approximately by 41 orders of magnitude. Further,  $\tilde{\sigma}_{\rm R_{\rm th}}\propto R^{-2}_{\rm th}$ for $R_{\rm th} \gg 1$  as expected from the behavior of sources near to the caustics \citep{2002glml.book.....M,CaminhaMagBias}.

For the PNFW model we have
\begin{equation}
\mathcal{S}(\phi)=x^2\left[a_{\varphi}  b_{\varphi} \mathcal{S}^{2}_{1\varphi}-(a_{\varphi} +b_{\varphi} )\mathcal{S}_{1\varphi}+\mathcal{S}_{0\varphi}\right], \label{S_pnfw}
\end{equation}
where
\begin{eqnarray*}
\mathcal{S}_{1\varphi}&=&\kappa(x_\varphi) +\kappa_s^\varphi, \\
\mathcal{S}_{0\varphi}&=&1 +(a_{\varphi} -b_{\varphi} )\kappa_s^\varphi\cos{2\phi_\varphi}.
\end{eqnarray*}
In this case, this function must be evaluated at $x_t$ and $x_{\pm}$ given in Eqs. (\ref{xt_pnfw}) and (\ref{cdc_pnfw}), respectively. 

For the ENFW model we have 
\begin{equation}
\mathcal{S}(\phi)=x^2\left[(a_\Sigma b_\Sigma)^2(\mathcal{J}_0\mathcal{J}_1) \mathcal{S}^2_{1\Sigma}-a_\Sigma b_\Sigma(\mathcal{J}_1\mathcal{P}+\mathcal{J}_0\mathcal{Q})\mathcal{S}_{1\Sigma}+\mathcal{S}_{0\Sigma}\right], \label{S_enfw}
\end{equation}
with
\begin{eqnarray*}
\mathcal{S}_{1\Sigma} &=&\kappa(x)+\kappa_s^\Sigma,\\
\mathcal{S}_{0\Sigma}&=&(a_\Sigma b_\Sigma)^2\mathcal{J}_0\mathcal{J}_1(\kappa_s^\Sigma)^2+\mathcal{PQ}-\gamma^2_{2\Sigma}.
\end{eqnarray*}
This function must be evaluated at $x_t$, Eq. (\ref{xt_enfw}), and $x_{\pm}$ Eq. (\ref{cdc_enfw}).

\section{Domain of validity of the solutions \label{val_sols}}
In this section we quantify the deviation of the analytic solutions (Sects. \ref{approx_nfws} and \ref{dcs_sects}) with respect to their corresponding exact calculations, seeking to determine their domains of validity in terms of the model parameters (Sect. \ref{lim_cdc}). Then, aiming to test the domain of validity of these solutions for computing  other lensing quantities, we compare the deformation cross sections in Sect. \ref{compar_dcs}.
\begin{figure}[!ht]
\begin{center}
\sidecaption \resizebox{\hsize}{!}{\includegraphics{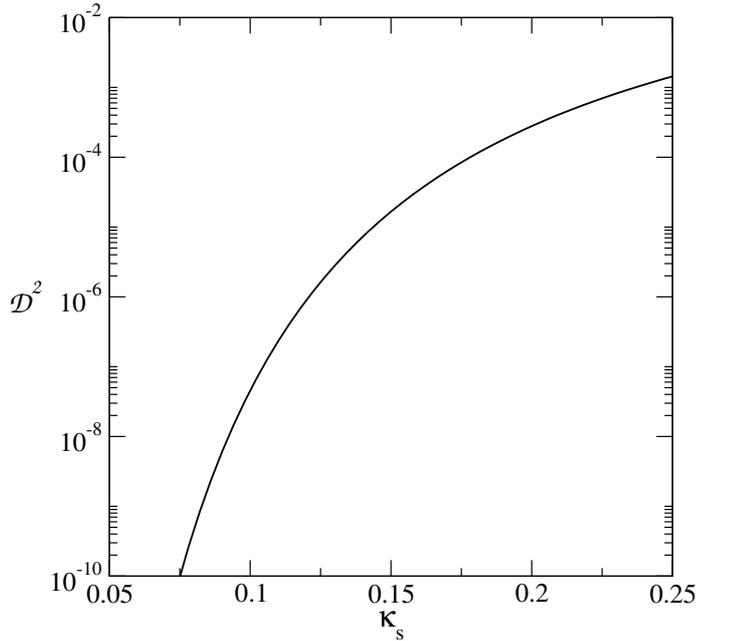}}
\caption{\label{fig1} Mean weighted squared radial fractional difference $\mathcal{D}^2$ for the constant distortion curve with $R_{\rm th} = 5$ for the circular NFW lens model as a function of the characteristic convergence $\kappa_s$.}
\end{center}
\end{figure}

\subsection{Limits for constant distortion curves and critical curves\label{lim_cdc} }
To compare the analytic solutions for critical curves and constant distortion curves to their exact calculations, we use as a figure-of-merit the mean weighted squared fractional difference between the two curves \citep{2013arXiv1301.0060D}
\begin{equation}
\mathcal{D}^2 = \frac{\sum_{i=1}^N w_i\left[x_{\rm ex}(\phi_i) - x_{\rm app}(\phi_i) \right]^2}{\sum_{i=1}^N w_i x^2_{\rm ex}(\phi_i)}, \label{D2_funct}
\end{equation}
where $N$ is the number of points of the curves, $\phi_i$ is their polar angle, $w_i = \phi_i-\phi_{i-1}$ is a weight (to  account for a possible non-uniform distribution of points), $x_{\rm ex}(\phi_i)$  and  $x_{\rm app}(\phi_i)$ are the radial coordinates of the curves curves obtained from the exact and approximated calculations, respectively. We notice that, the $\mathcal{D}^2$ in Eq. (\ref{D2_funct}) is independent both of the lens length scale and of the discretization.
\begin{figure*}[!ht]
\centering \sidecaption \resizebox{\hsize}{!}{
\subfigure{\includegraphics{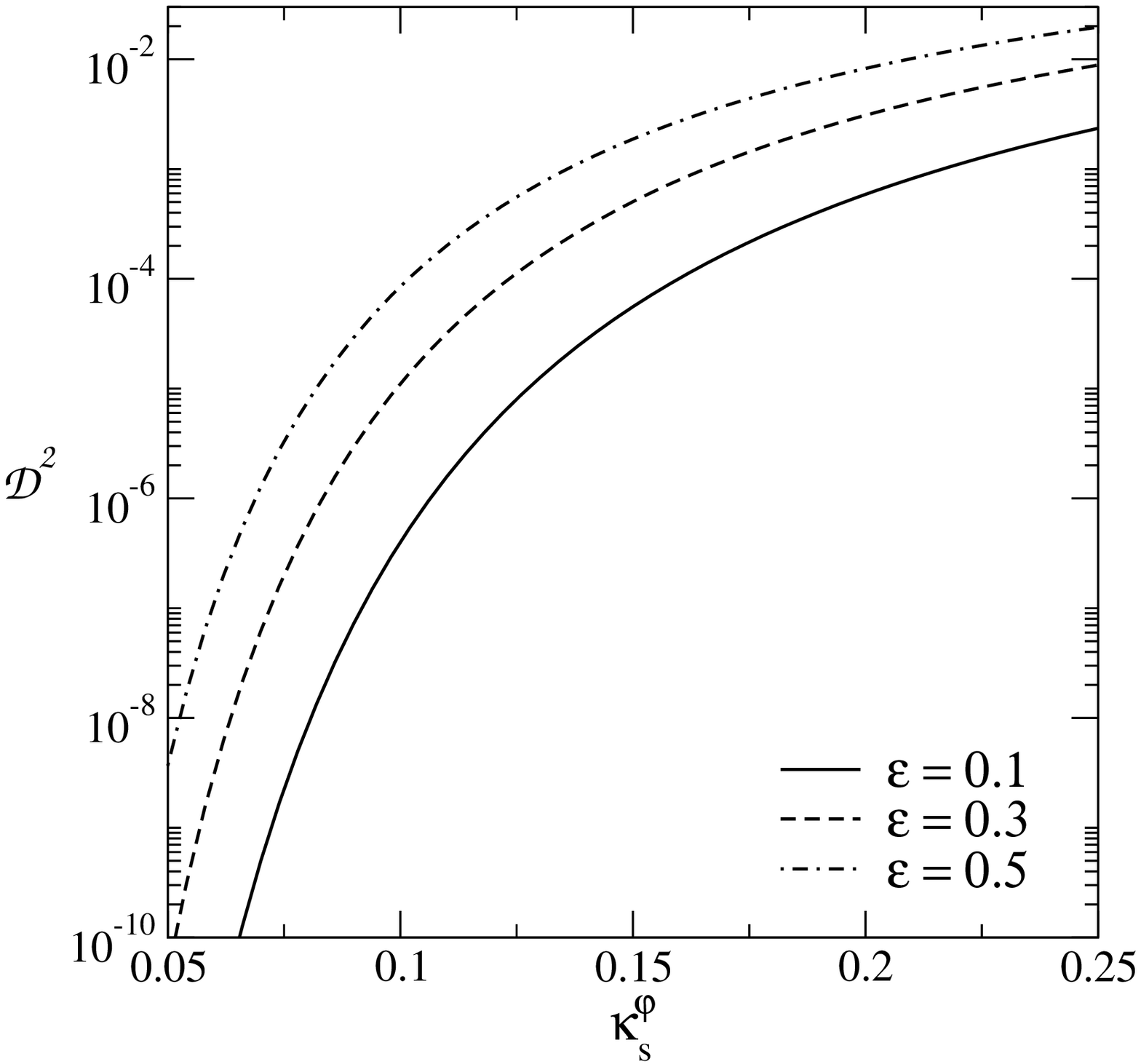}} \hspace{1.5cm}
\subfigure{\includegraphics{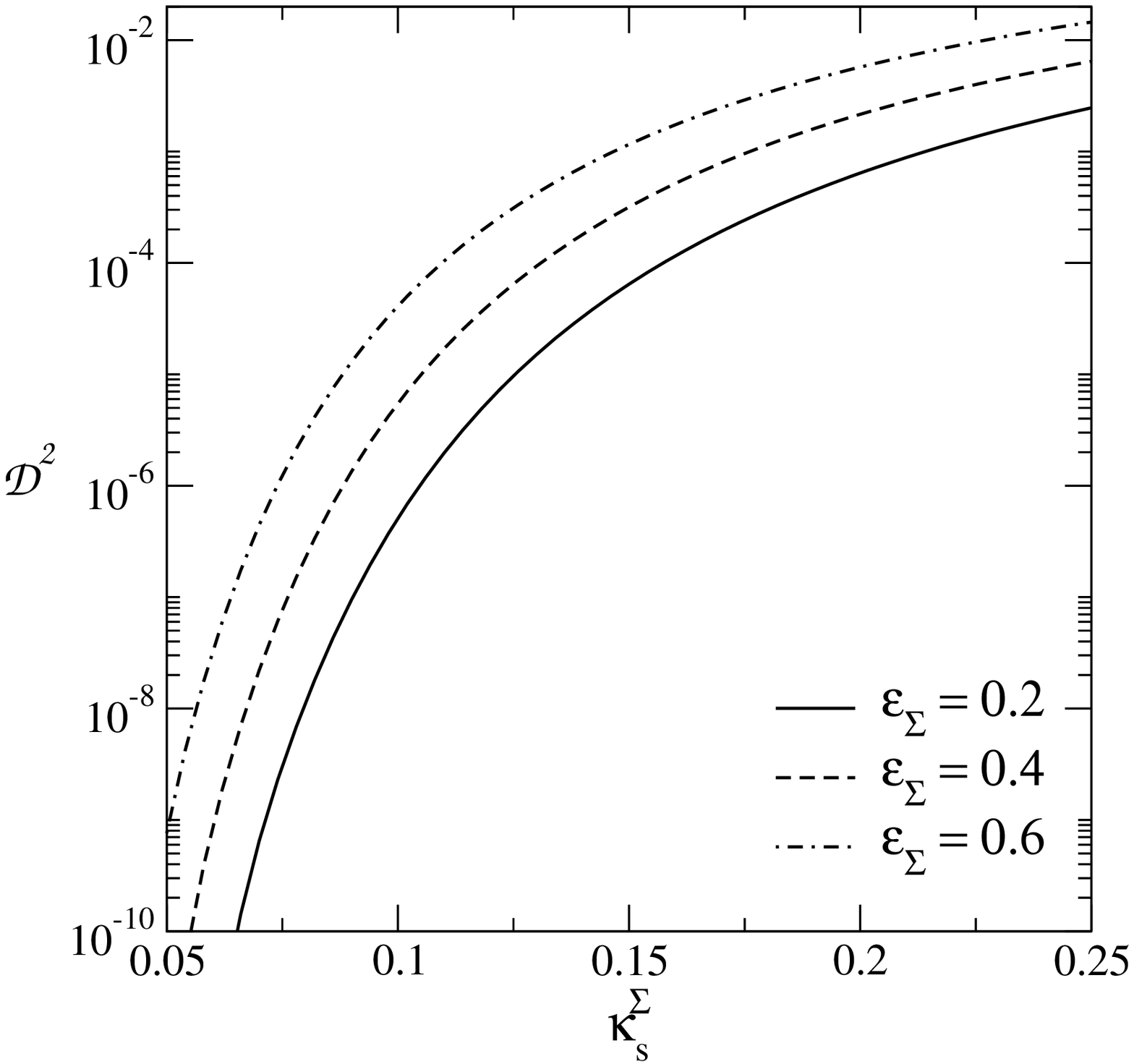}}}
\caption{\label{Fig2}  Mean weighted squared radial fractional difference $\mathcal{D}^2$, for constant distortion curves with $R_{\rm th}=5$, as a function of the characteristic convergences, for some values of ellipticity parameters. Left panel: PNFW model. Right panel: ENFW model.}
\end{figure*}

Choosing a maximum  value for $\mathcal{D}^2$, we can define upper values for the characteristic convergences (for a given ellipticity) such that the contour curves obtained with both the exact and approximated calculations will be close enough to each other. This maximum value is chosen by visually comparing the approximated and exact solutions for critical curves and constant distortion curves for several values of $\mathcal{D}^2$ and combinations of the NFW model parameters.

We compute $\mathcal{D}^2$ on the $R_\lambda=+R_{\rm th}$ curves since their points are the farthest from the lens center.
Thus, imposing a limit on the NFW parameters to match the $R_\lambda=R_{\rm th}$ curve, we automatically match the curves enclosed by it. We have checked that for a broad range of the NFW model parameters and ellipticities (for both elliptical and pseudo-elliptical models), a good visual matching of the exact and approximated curves if obtained for a maximum value of  $\mathcal{D}^2 = 10^{-4}$ (for $R_{\rm th}=5$). This value also ensures a good matching of the critical curves, and we found that the corresponding curves in the source plane are well-matched, too. We thus fix this as the upper value of  $\mathcal{D}^2$ throughout this work.

In Fig. \ref{fig1} we show $\mathcal{D}^2$ for the $R_{\rm th}=5$ curve as a function of $\kappa_s$ for the circular NFW model. Setting the above upper value for  $\mathcal{D}^2$ leads to a maximum value of $\kappa_s = 0.18$, for which all curves $|R_\lambda|= R_{\rm th}(\geq 5)$, and the radial critical curves are well matched. 
\begin{figure*}[!htp]
\resizebox{0.98\hsize}{!}{
\subfigure{\includegraphics{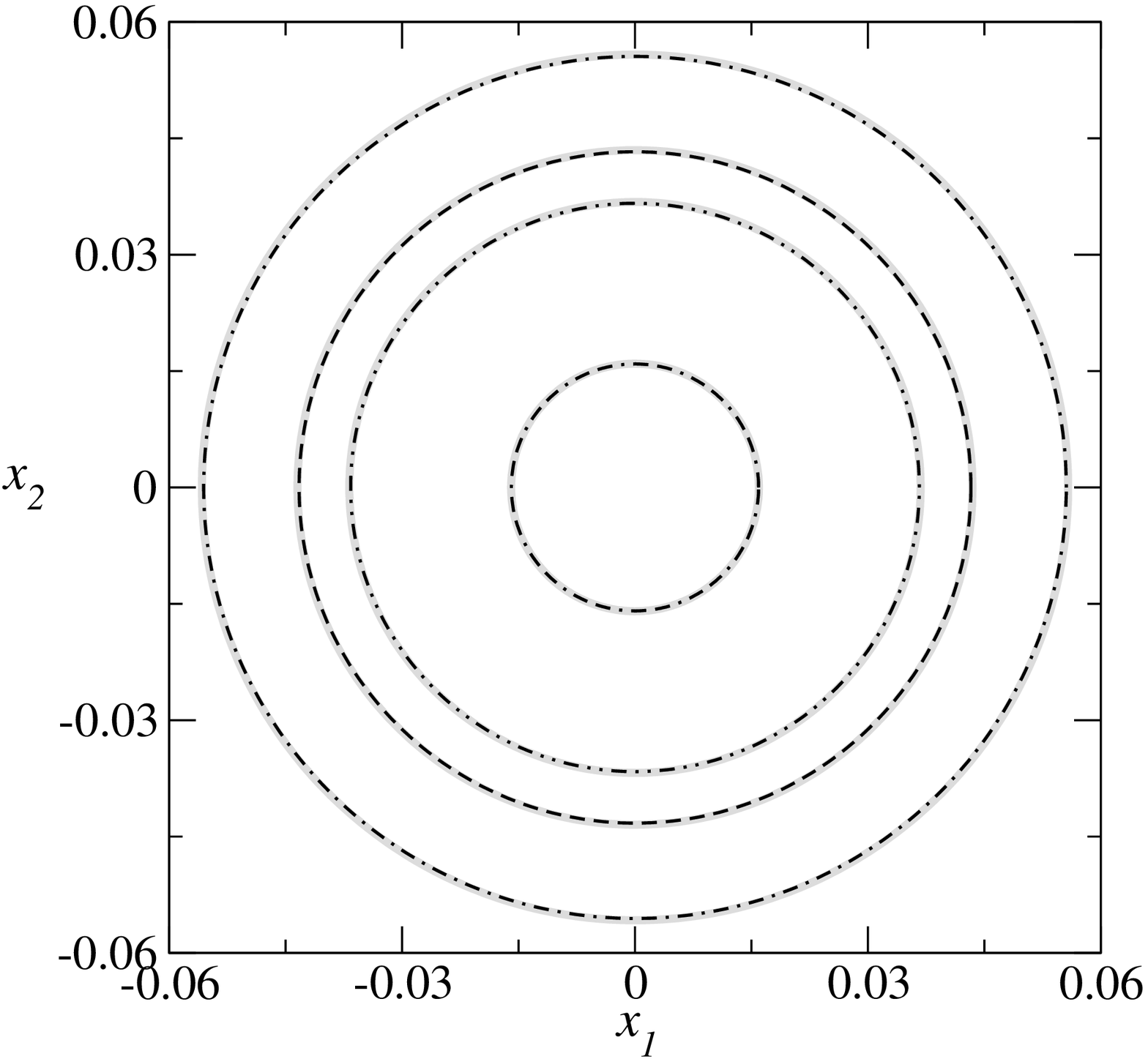}} \hspace{2.cm}
\subfigure{\includegraphics{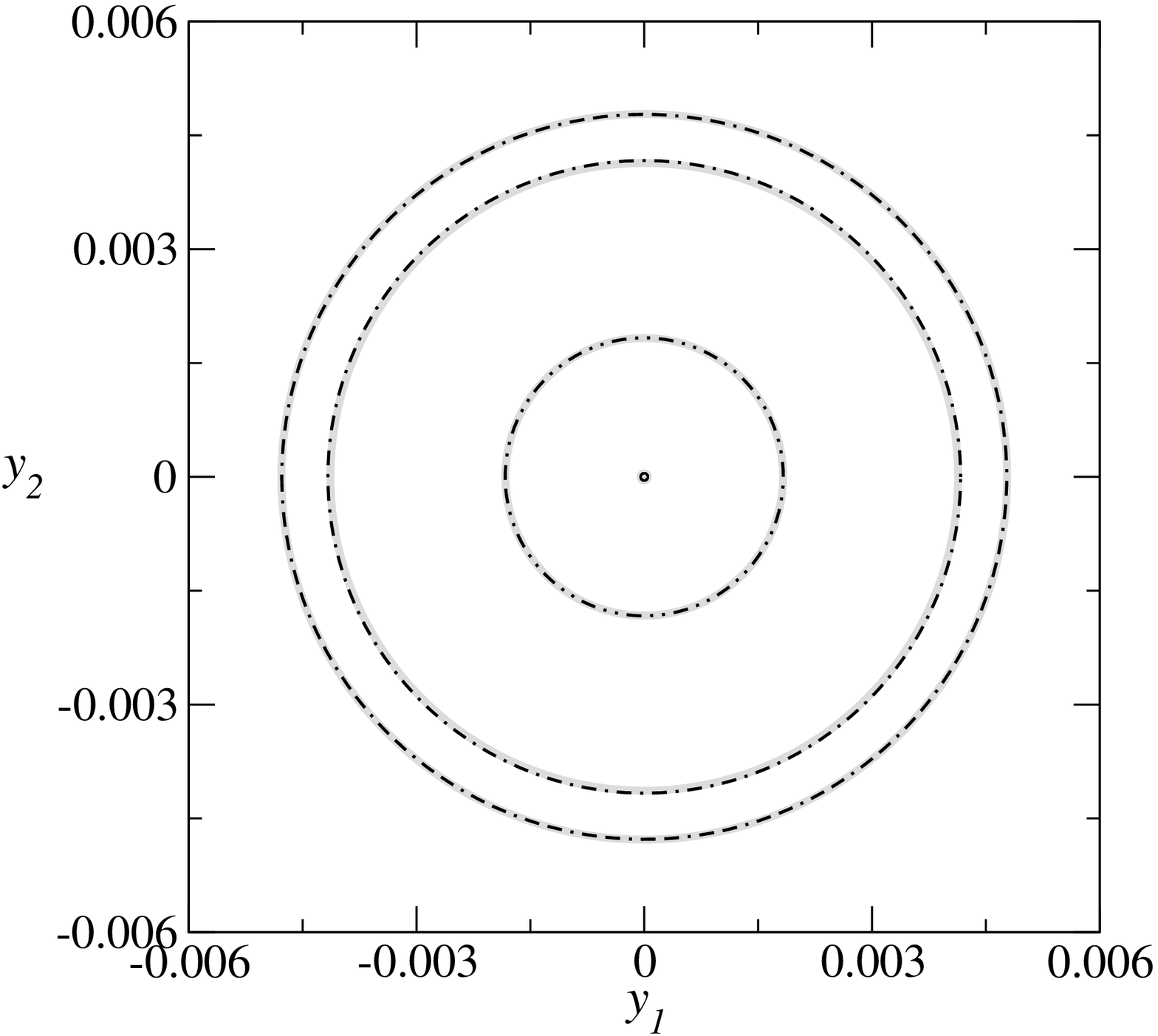}}}

\resizebox{0.98\hsize}{!}{
\subfigure{\includegraphics{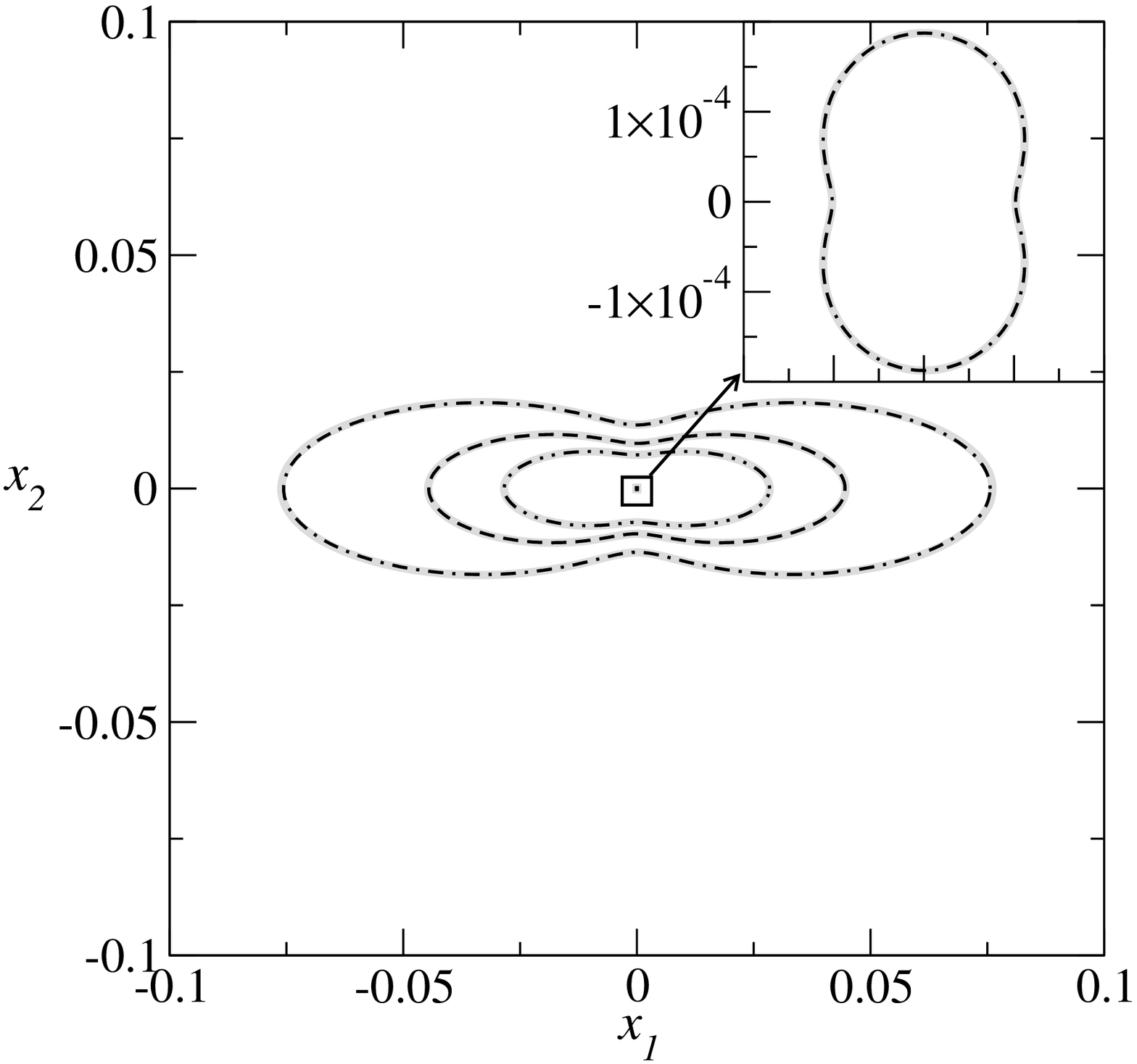}} \hspace{2.cm}
\subfigure{\includegraphics{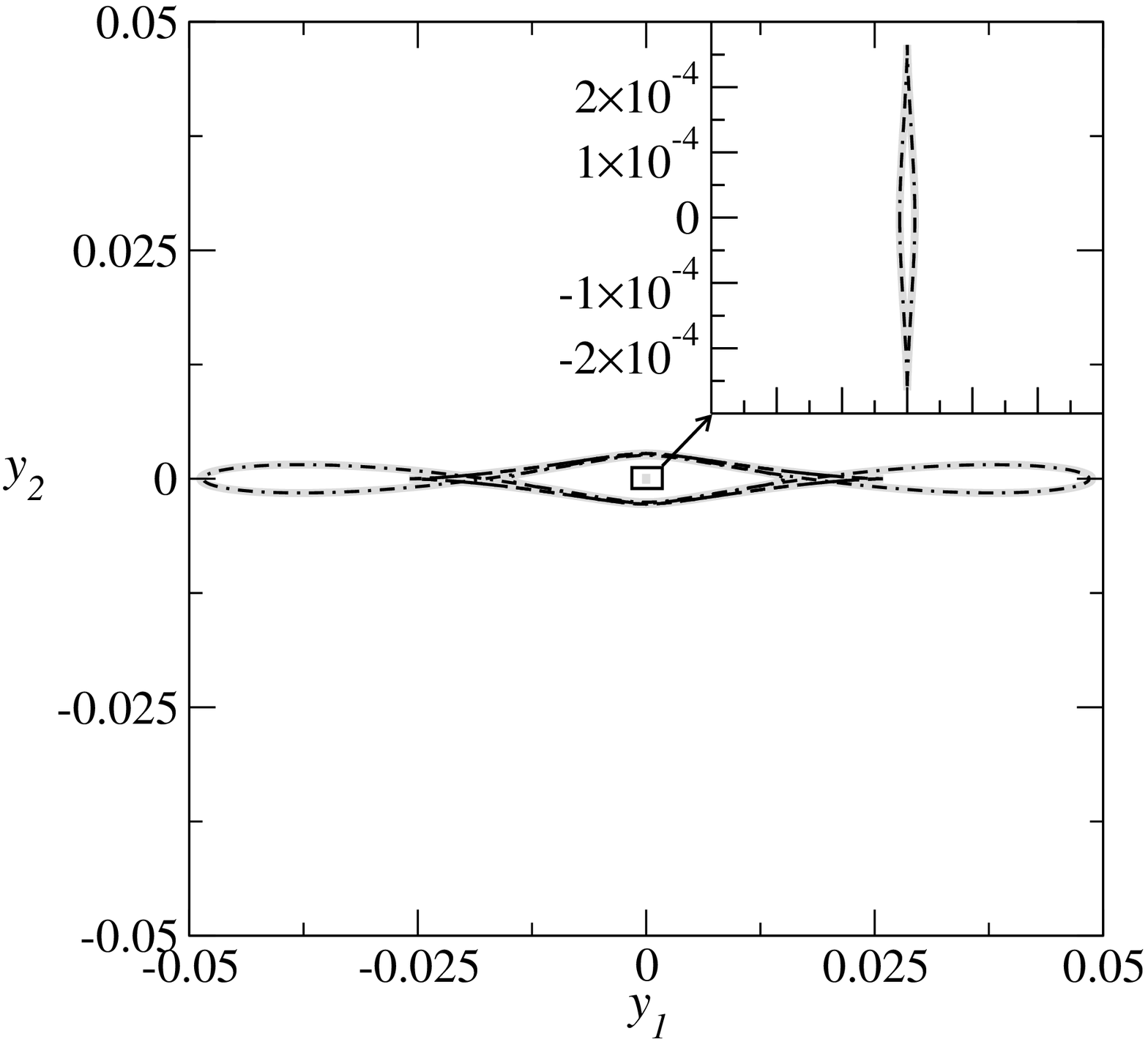}}}

\resizebox{0.98\hsize}{!}{
\subfigure{\includegraphics{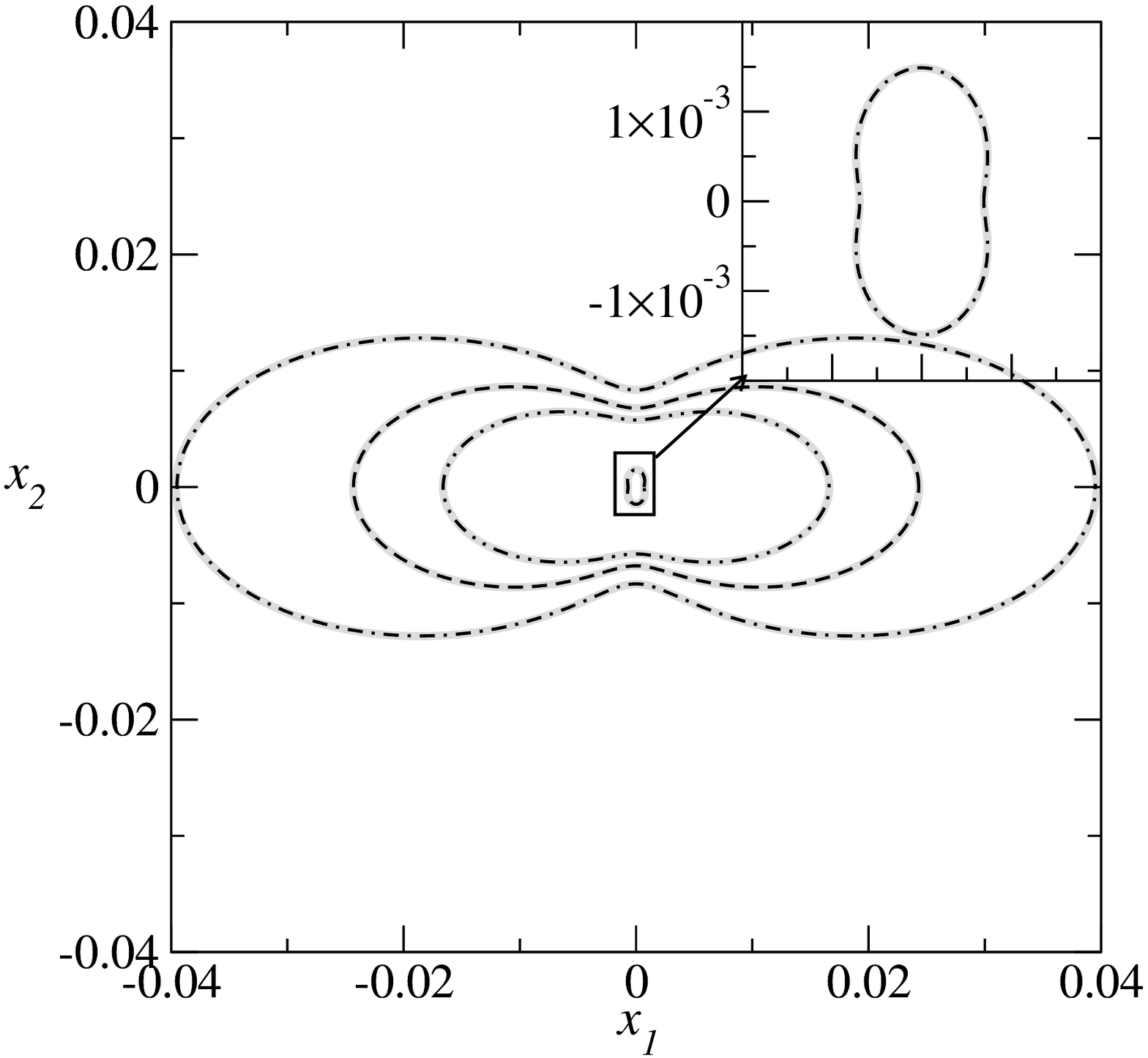}} \hspace{2.cm}
\subfigure{\includegraphics{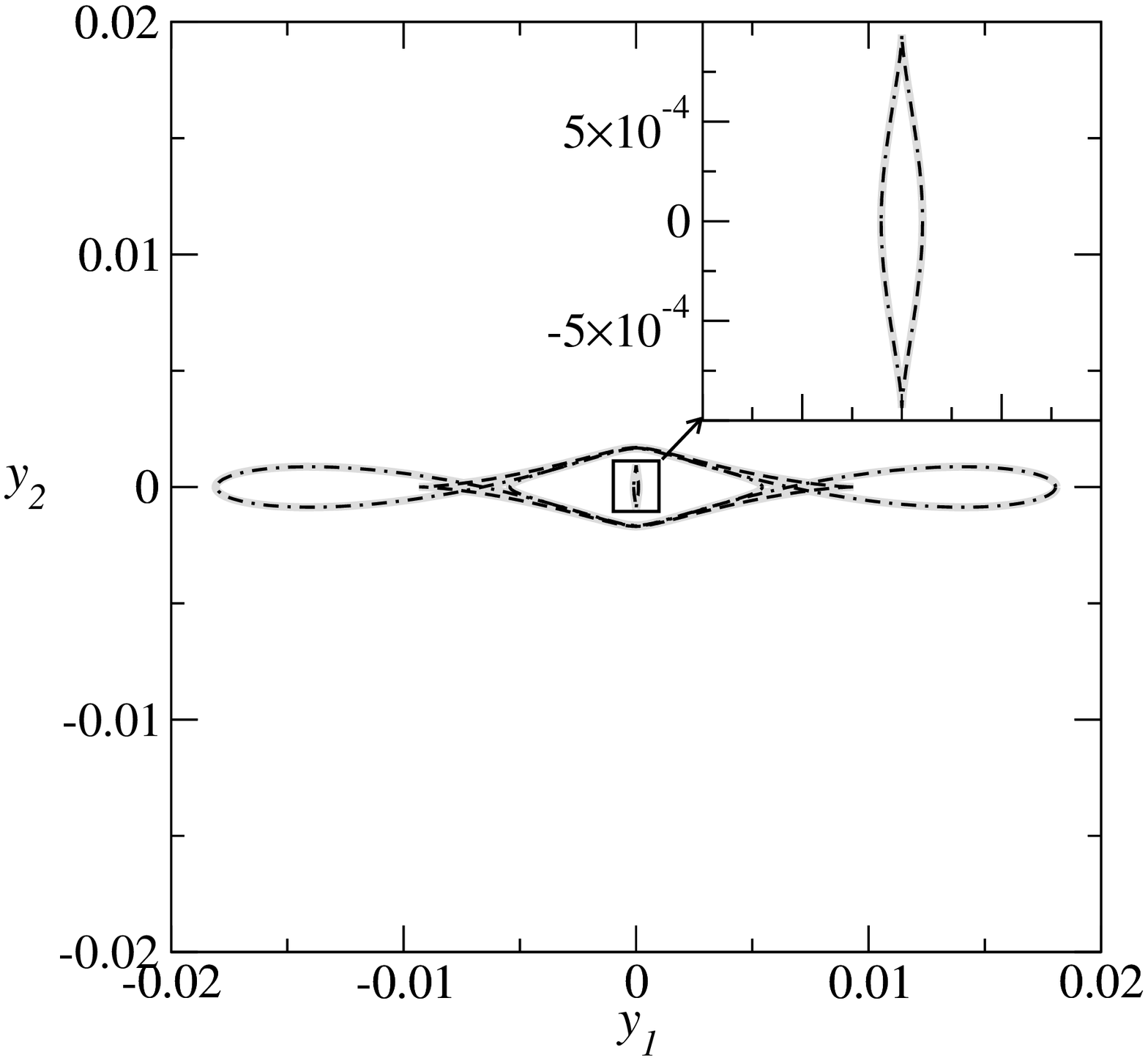}}}
\caption{\label{Fig3} Constant distortion curves $R_\lambda = +R_{\rm th}$ (dot-dashed lines), $R_\lambda = -R_{\rm th}$  (dot-dot-dashed lines) for $R_{\rm th}=5$, tangential curves (dashed lines), and radial curves (dash-dash-dotted lines) obtained with the exact (thick gray lines) and approximated (black lines) calculations in the lens plane (left panels) and source plane (right panels). Upper panels: NFW model for $\kappa_s=0.15$. Middle panel: PNFW model for $\kappa^\varphi_s=0.1$ and $\varepsilon=0.5$. Bottom panels: ENFW model for $\kappa^\Sigma_s=0.1$ and $\varepsilon_\Sigma=0.4$. In the right panels the axes are in units of $\eta_0=r_s D_{\rm OS}/D_{\rm OL}$.}
\end{figure*}

For the PNFW model, we chose the convention \citep{1987ApJ...321..658B,2002A&A...390..821G}
\begin{equation}
 a_{\varphi}=1-\varepsilon, \quad b_{\varphi}=1+\varepsilon,
\label{choice_gk02}
\end{equation}
where the ellipticity parameter $\varepsilon$ is defined in the range $ 0 \leq \varepsilon < 1$.
As shown in DCM, for low values of $\kappa^\varphi_s$ we must have $\varepsilon \lesssim 0.5$ to avoid  dumbbell-shape mass distributions. In the forthcoming analyses, we therefore adopt an upper value of $\varepsilon=0.5$.
In the left hand panel of Fig. \ref{Fig2} we show $\mathcal{D}^2$ as a function of $\kappa^\varphi_s$ for some values of $\varepsilon$. 
For the higher value of $\varepsilon$ considered, $\mathcal{D}^2 = 10^{-4}$ is reached for 
$\kappa_s^\varphi=0.1$, providing an upper limit for the characteristic convergence for the validity of the approximations. However, higher values of $\kappa^\varphi_s$ are allowed as $\varepsilon$ decreases. For instance, when $\varepsilon \rightarrow 0$, we find that $\kappa^\varphi_s \rightarrow 0.18$ (in agreement with the result for  the circular NFW model). 

For the ENFW model, we chose the parametrization
\begin{equation}
a_\Sigma=  \frac{1}{\sqrt{1-\varepsilon_{\Sigma}}} \ \ \mathrm{and} \ \ b_\Sigma =  \sqrt{1-\varepsilon_{\Sigma}},
\end{equation}
where the ellipticity $\varepsilon_\Sigma$ is defined in the range $ 0 \leq \varepsilon_\Sigma < 1$.
In the right hand panel of Fig. \ref{Fig2} we show $\mathcal{D}^2$ as a function of $\kappa^\Sigma_s$ for some values of $\varepsilon_\Sigma$. The behavior of $\mathcal{D}^2$, as well as the maximum values of $\varepsilon_\Sigma$ for each $\kappa_s^\Sigma$, is qualitatively similar to that of the PNFW model. In this case, we find that 
a maximum value of $\kappa^\Sigma_s=0.1$ is allowed for $\varepsilon_\Sigma=0.7$. Again the maximum value  $\kappa^\Sigma_s \rightarrow 0.18$ as $\varepsilon_\Sigma \rightarrow 0$.
 
In Appendix \ref{app_fit} we present fitting functions for the maximum values of $\varepsilon$ ($\varepsilon_\Sigma$) 
as a function of $\kappa^\varphi_s$ ($\kappa^\Sigma_s$) for which the approximation is accurate following the criteria of this section. In Fig. \ref{Fig3} we show the constant distortion curves (including tangential and radial critical curves) in both the lens and source planes, for some values of the NFW, PNFW, and ENFW parameters, both using the approximations introduced in this work, as well as through the numerical computation with no approximations. The parameters were chosen such that $\mathcal{D}^2 \lesssim 2\times 10^{-5}$, i.e. less than our cut $\mathcal{D}^2=10^{-4}$. We see that the exact and approximated curves are almost indistinguishable, 
visually illustrating the validity of using $\mathcal{D}^2$ to determine the domain of validity of the approximations.

\subsection{Comparison between deformation cross sections \label{compar_dcs}}

We compare the exact calculations and the solutions derived in Sect. \ref{dcs_subsect}, in order to verify that the limits derived in Sect. \ref{lim_cdc} also give a domain of validity for the deformation cross section.
To quantify this comparison, we compute the relative difference
\begin{equation}
\frac{\Delta \tilde{\sigma}_{R_{\rm th}}}{\tilde{\sigma}_{R_{\rm th}}}=\left|\frac{\tilde{\sigma}_{R_{\rm th},\rm  ex}-\tilde{\sigma}_{R_{\rm th},\rm app}}{\tilde{\sigma}_{R_{\rm th},\rm ex}}
\right|,\label{dif_rela_sigma}
\end{equation}
where the subscripts \emph{ex} and \emph{app} refer to the exact and approximated calculations, respectively.

\begin{figure}[!htb]
\begin{center}
\sidecaption \resizebox{\hsize}{!}{\includegraphics{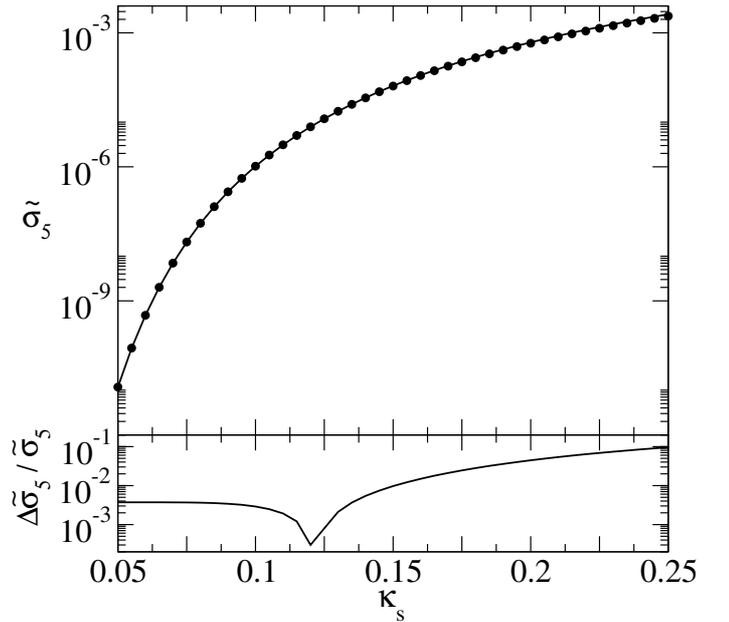}}
\caption{ \label{Fig4} Deformation cross section for the circular NFW lens model as a function of $\kappa_s$. Solid line corresponds to expression (\ref{dcs_nfw_analytic}). Filled circles correspond to the exact (numerical) calculation.}
\end{center}
\end{figure}

\begin{figure*}[!ht]
\centering \sidecaption \resizebox{\hsize}{!}{
\subfigure{\includegraphics{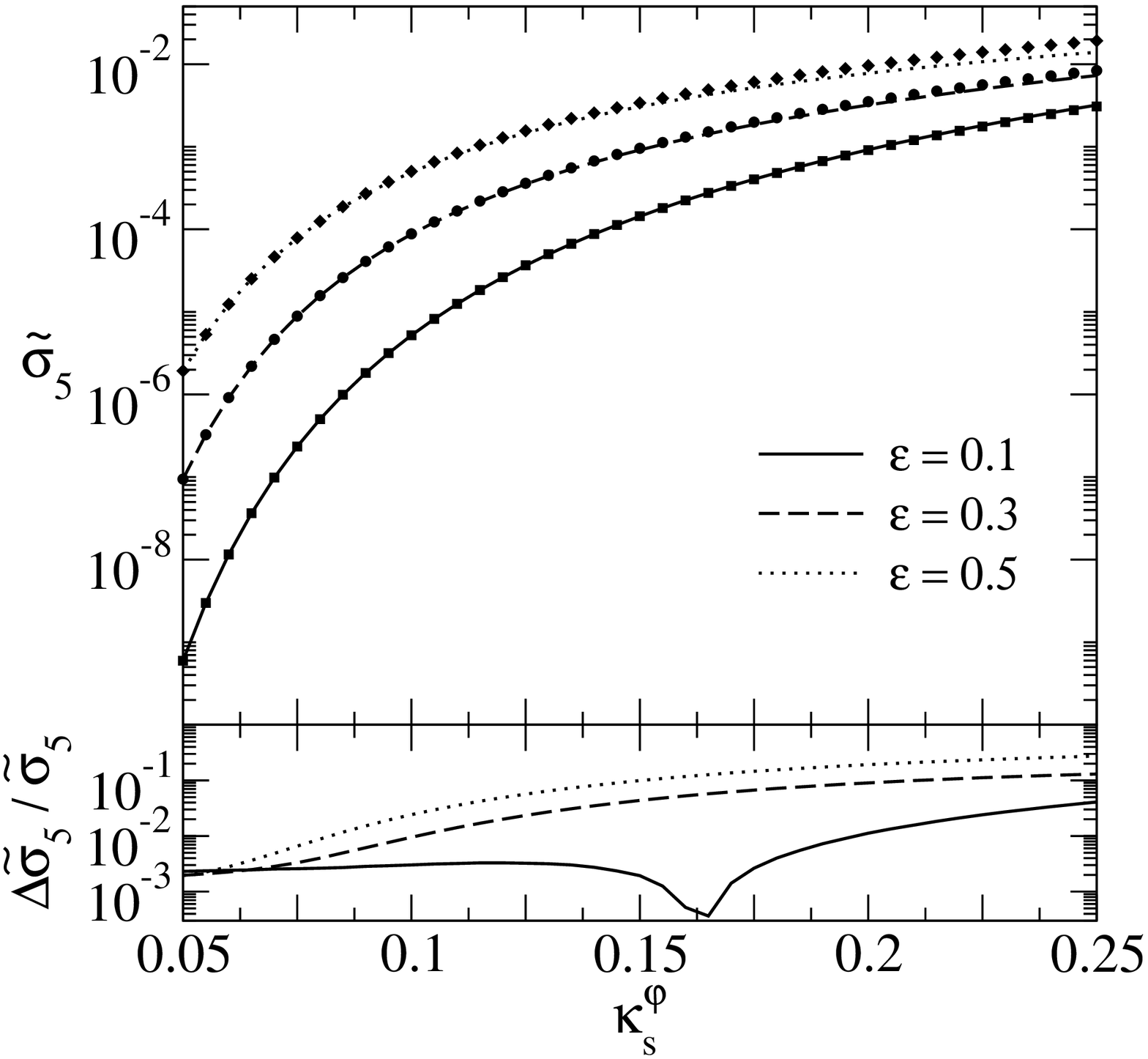}} \hspace{1.cm}
\subfigure{\includegraphics{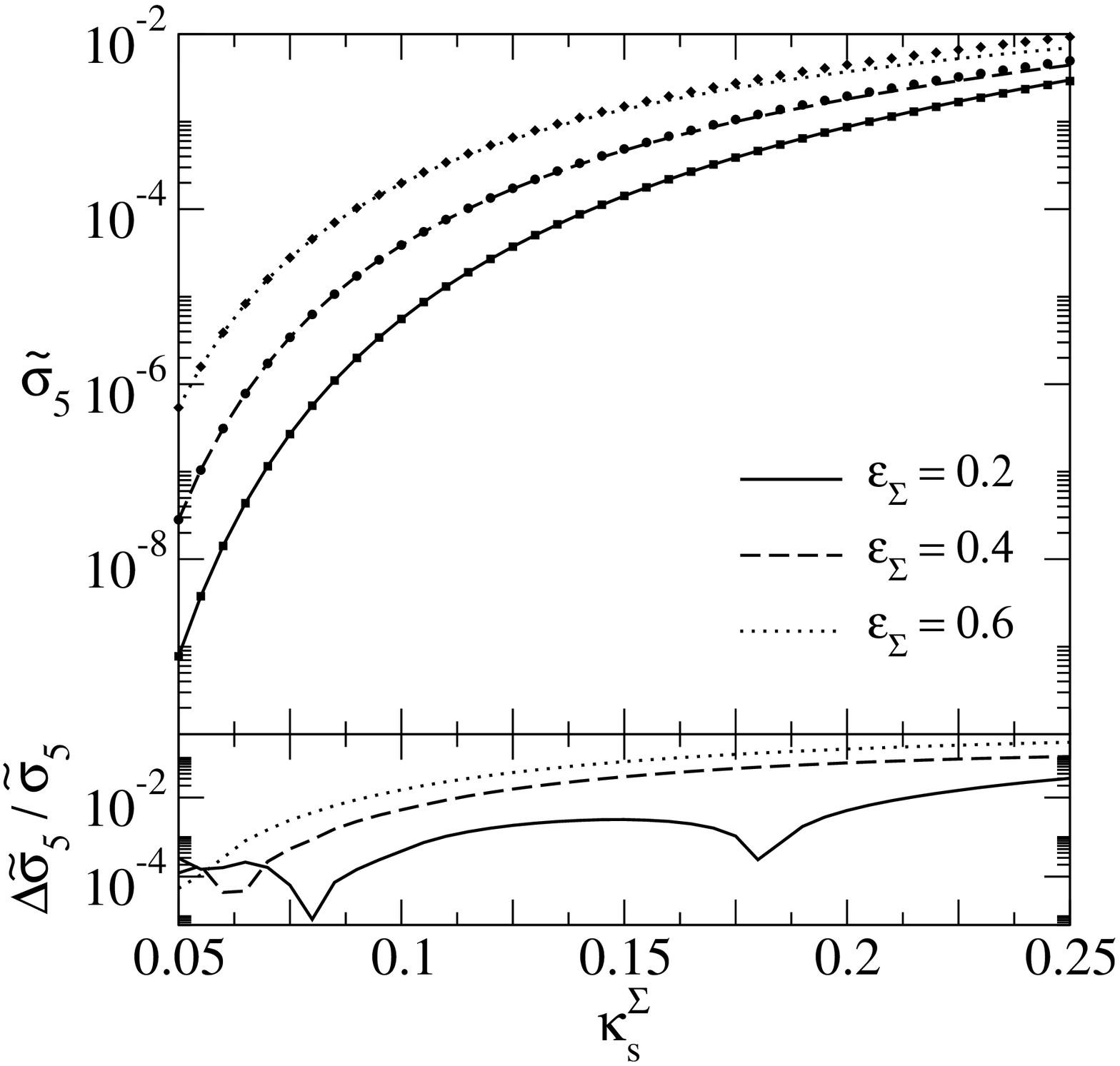}}}\caption{\label{Fig5} Deformation cross sections for the PNFW model (left panel) and ENFW model (right panel). Symbols correspond to the exact calculation.} 
\end{figure*}

In Fig. \ref{Fig4} we show $\tilde{\sigma}_{5}$ (upper panel) and $\Delta \tilde{\sigma}_{5}/\tilde{\sigma}_{5}$ (bottom panel) as a function of $\kappa_s$ for both the exact and approximated calculations. Considering the limit of the previous section, i.e., $\kappa_s \leq 0.18$, Eq. (\ref{dcs_nfw_analytic}) deviates from the exact calculation by at most 2.5\%. 

In the left hand panel of Fig. \ref{Fig5} we show $\tilde{\sigma}_{5}$ (upper panel) and $\Delta \tilde{\sigma}_{5}/\tilde{\sigma}_{5}$ (bottom panel) as a function of $\kappa^\varphi_s$ for the PNFW model for both the exact and the approximated calculations. 
Considering the limit of applicability obtained in the previous section ($0.1 \leq \kappa^\varphi_s \leq 0.18$ and $\varepsilon\leq 0.5$),  Eq. (\ref{sigma_medio}), with Eq. (\ref{S_pnfw}), deviates at most 2.6\% from the exact calculation. 
A very similar result is found for the ENFW model. In this case, for  $ 0.1\leq \kappa_s^\Sigma \leq 0.18$ and $\varepsilon_\Sigma \leq  0.7$, we find that $\Delta \tilde{\sigma}_{5}/\tilde{\sigma}_{5}$ ($\tilde{\sigma}_{5}$ calculated by Eq. (\ref{sigma_medio}) with Eq. (\ref{S_enfw})) is at most $2.5\%$.

\section{Mapping among PNFW and ENFW parameters \label{mapping}}

As mentioned in the introduction, elliptical models are more physically motivated than pseudo-elliptical ones. However, 
determination of lensing quantities, such as the shear,  for the former requires evaluating integrals \citep[][see also Appendix \ref{app_enfw}]{1990A&A...231...19S,2001astro.ph..2341K}, which generally have to be computed numerically.
On the other hand, pseudo-elliptical models do not require such integrals to be evaluated (see, e.g., Eqs. (\ref{aphi})--(\ref{g2_pnfw})) allowing for fast calculations for the same quantities. 
However, it is well known that pseudo-elliptical models have two main limitations: their surface mass density can assume negative values in some regions \citep{1987ApJ...321..658B} and may present a ``dumbbell'' shape for high ellipticities \citep{1989ApJ...337..621K,SEF,1993ApJ...417..450K}. At least in the case of the NFW model, the first problem does not affect the region of arc formation (DCM). Regarding the shape of the iso-convergence contours, for each value of $\kappa_s^\varphi$, there is a range in $\varepsilon$ for which the contours are approximately elliptical (DCM). In principle, within this range of parameters, the pseudo-elliptical models could be employed instead of elliptical ones in studies that require numerous evaluations of lensing quantities. For this sake, a correspondence among model parameters has to be established, which associates a pair of the PNFW parameters ($\varepsilon$, $\kappa_s^{\varphi}$) to a pair of the ENFW ones ($\varepsilon_{\Sigma}$, $\kappa_s^{\Sigma}$).

To obtain $\varepsilon_{\Sigma}$ from the PNFW model we use the same procedure as in \citet[][see also DCM]{2002A&A...390..821G}. From Eq. (\ref{xk_pnfw}) we define the semi-major axis $a$ by $x_\kappa(\phi=0)$ and the semi-minor axis $b$ by $x_\kappa(\phi=\pi/2)$ such that 
\begin{eqnarray}
\varepsilon_\Sigma = 1-\frac{b}{a} & = & 1-\sqrt{\frac{a_{\varphi} }{b_{\varphi} }}\exp{\left[\frac{(a_{\varphi} -b_{\varphi} )}{(a_{\varphi} +b_{\varphi} )} \right]}, \nonumber \\
    & =& 1-\sqrt{\frac{1-\varepsilon}{1+\varepsilon}}\exp{\left[-\frac{\varepsilon}{2} \right]}. \label{ellip_gk}
\end{eqnarray}
This expression  has no dependence on $R_{\rm th}$ or on $\kappa^\varphi_s$.

The qualitative behavior of $\varepsilon_\Sigma(\varepsilon)$ from this equation is very similar to what was found numerically for generic values of  $\kappa^\varphi_s$ in DCM (see, e.g., their Fig. 6). 
 For $\varepsilon \ll 1$, the expression above  gives  $\varepsilon_\Sigma \simeq 2\varepsilon$, in agreement with DCM (see,  e.g., Eq. (B.2), which gives $\varepsilon_\Sigma=1.97\varepsilon$, for $\kappa^\varphi_s \rightarrow 0$).   
 
Interestingly, Eq.  (\ref{ellip_gk}) is a good approximation of the exact relation for values of  $\kappa^\varphi_s$ well above the limit of validity of the approximation derived in Sect. \ref{val_sols}.
Indeed, the relative deviation with respect to the exact calculation of $\varepsilon_\Sigma(\varepsilon,\kappa^\varphi_s, R_{\rm th})$ (see the elliptical fit method in Sect. 4 of DCM) is at most  5\% for $\kappa^\varphi_s \leq 0.4$, with
$\varepsilon \leq 0.5$  and $R_{\rm th}\geq 5$. 

To associate a value of $\kappa^\Sigma_s$ to a pair ($\kappa^\varphi_s$, $\varepsilon$), we require that  the tangential critical curves of the PNFW and ENFW models match 
at $\phi=0$.
This condition, from Eqs. (\ref{xt_pnfw}) and (\ref{xt_enfw}), gives
\begin{equation}
\kappa_s^\Sigma=\frac{b_{\varphi} \kappa_s^\varphi}{a_\Sigma b_\Sigma\left[\mathcal{J}_1%
+b_{\varphi} \kappa^\varphi_s(\mathcal{J}_1\log{a_{\varphi} }-\mathcal{J}_1-2\mathcal{L}_1(0))\right]}.
\label{ka_rela_gk}
\end{equation}
For $\varepsilon =\varepsilon_\Sigma =0$ the expression above reduces to $\kappa_s^\varphi=\kappa_s^\Sigma=\kappa_s$, as it should be. 

We determine the upper value of $\kappa^\varphi_s$ such that the expression above deviates by at most 5\% with respect to the exact calculation. This yields $\kappa^\varphi_s < 0.5$ for $\varepsilon \leq 0.5$ and R$_{\rm th} \geq 5$.
Again, the domain of validity of the approximate mapping is greater than for the lensing equations.

\section{Summary and concluding remarks  \label{s&c}}

When considering low values of the characteristic convergence of the NFW model (i.e., on the galactic and galaxy group mass scales), some strong lensing quantities require high numerical precision to yield accurate results, which can demand a lot of time. Motivated by this issue, we obtained analytic solutions for several strong lensing quantities for elliptical and pseudo-elliptical NFW models and quantified their corresponding limits of validity.

The starting point is approximation (\ref{angle_nfw_tot}) for the  deflection angle of the circular NFW model. This approximation was applied to the standard prescriptions for obtaining the convergence and shear for circular, elliptical, and pseudo-elliptical models, leading to analytic solutions for these quantities (Sect. \ref{approx_nfws}). Those were in turn used to derive analytic expressions for iso-convergence contours and critical curves (see Sect. \ref{approx_nfws}) and for the constant distortion curves as a function of $R_{\rm th}$ (Sect. \ref{cdc_nfws}).
 
As a practical application of these results, we computed the deformation cross section ($\sigma_{\rm R_{\rm th}}$).
In the case of the circular NFW model, we obtained an analytical formula for $\sigma_{\rm R_{\rm th}}$ (Eq. (\ref{dcs_nfw_analytic})), which reproduces the behavior 
with respect to $\kappa_s$ and the scaling with $R_{\rm th}$ obtained numerically in previous works  \citep[see, e.g.,][]{CaminhaMagBias}. 
We have shown that the computation of this cross section is reduced to a one-dimensional integral for both the PNFW  and ENFW models (Eq. (\ref{sigma_medio}) with either Eq. (\ref{S_pnfw}) or Eq. (\ref{S_enfw}), respectively). 
These expressions speed up the numerical computations by two orders of magnitude for the PNFW model and one order of magnitude for the ENFW one, independently of the values of the ellipticity parameter.

We used the figure-of-merit $\mathcal{D}^2$, Eq. (\ref{D2_funct}), to quantify the deviation of the solutions of the constant distortion curves with respect to their exact calculations. Setting a maximum value of $\mathcal{D}^2 = 10^{-4}$,  we find that Eq. (\ref{cdc_pnfw}) (Eq. (\ref{cdc_enfw})), matches its corresponding exact calculation up to $\kappa^\varphi_s =0.1$ ($\kappa^\Sigma_s =0.1$) for $\varepsilon\leq 0.5$ ($\varepsilon_\Sigma \leq 0.7$) and $R_{\rm th}>5$. In particular, we find that the characteristic convergence can go up to $0.18$ as the ellipticity parameters tend to zero, as expected from the limit derived for the circular NFW model. 
In Appendix \ref{app_fit} we provide fitting functions for the maximum ellipticity allowed as a function of the characteristic convergences (in the range $0.1 <\kappa^\Sigma_s,\kappa^{\varphi}_s  < 0.18$) to ensure the validity of the approximations.  We emphasize that these limits also  ensure a good match for critical curves and the  $R_\lambda = - R_{\rm th}$ curves to their corresponding exact calculations, since these curves 
 are enclosed by the $R_\lambda = R_{\rm th}$ curve. 
We verified that the corresponding curves in the source plane also match the exact solution for the parameters within the limits derived above.

We compared the deformation cross sections (obtained from both the exact and approximated calculations) in order to check that the domain of validity derived for the constant distortion curves also holds for this quantity (Sect. \ref{compar_dcs}), which we found to be the case. For instance, for the circular NFW model, for $\kappa_s \leq 0.18$ and $R_{\rm th}=5$, Eq. (\ref{dcs_nfw_analytic}) deviates at most 2.5\% from the exact calculation, while for the PNFW and ENFW, $\Delta\tilde{\sigma}_5/\tilde{\sigma}_5$ (Eq. (\ref{dif_rela_sigma})) is at most $2.5\%$, for  $0.1 \leq \kappa^\varphi_s \leq 0.18$ and $\varepsilon<0.5$, and  $ 0.1 \leq \kappa^\Sigma_s \leq 0.18$ and $\varepsilon_\Sigma < 0.7$, respectively.

Overall, the approximate solutions presented here are accurate for all strong lensing quantities that were addressed in this work, within the considered parameter ranges, for $\kappa_s \leq 0.1$. In some cases they are valid for higher  $\kappa_s$, such as for low ellitpcities. Furthermore, some derived quantities are valid up to much higher characteristic convergences. For example, the  ellipticity of the iso-convergence contours of the PNFW model is reproduced well by the simple analytic form of Eq.~(\ref{ellip_gk}) up to $\kappa^\varphi_s \simeq 0.4$. We found that in this range the relation $\varepsilon_\Sigma(\varepsilon)$ is independent of $\kappa^\varphi_s$ and of the chosen value of the contour. To complete the association of the PNFW to ENFW model parameters we derived a relation among characteristic convergences, Eq.  (\ref{ka_rela_gk}), which also matches the values obtained in DCM for  $\kappa^\varphi_s \le 0.5$ and $R_{\rm th}\ge5$.

The analytic solutions presented here allow for a robust and fast computation of several strong lensing quantities to be carried out in the low characteristic convergence regime. They may thus be useful for the lensing community 
and could be readily included in strong lensing codes, considering the domain of applicability derived in this work.

In principle, the approximate solutions derived in this paper could be extended to other quantities, such as the lensing magnification and the magnification cross section. They might also be applied to finding solutions of the lens equation, for multiple images and arcs. One way of finding approximate solutions for arcs, for low ellitpicities, is through the perturbative approach \citep{2007MNRAS.382L..58A,2008MNRAS.388..375A,2008MNRAS.390..945P}, for which the Einstein ring solution (given analytically in Eq. (\ref{xt_nfw})) is the starting point. 
Therefore, analytic solutions for arcs might be derived in this framework. The investigation of these and other possible extensions is left for future work.

\begin{acknowledgements}
We thank the anonymous referee for useful suggestions that helped improve this manuscript. We thank the anonymous referee of DCM for pushing us to consider the low $\kappa_s$ regime. H.~S.~D\'umet-Montoya is funded by the Brazilian agency CNPq. G.~B.~Caminha is funded by CNPq and CAPES.  M.~Makler is partially supported by CNPq (grant 309804/2012-4) and FAPERJ (grant E-26/110.516/2012). 
\end{acknowledgements}

\begin{appendix}
\section{ Lensing functions for elliptical lens models \label{app_enfw}}
The lensing functions of models with elliptical mass distribution can be obtained following 
the expressions derived in \citet{1990A&A...231...19S} and \citet{2001astro.ph..2341K}, which were generalized for any choice  of the ellipticity parameterization in  \citet{CaminhaMagBias} and are given by
\begin{eqnarray}
\alpha_{1\Sigma} & = & a_{\Sigma}b_{\Sigma}  J_0 x \cos{\phi}, \quad \alpha_{2\Sigma} = a_{\Sigma}b_{\Sigma}  J_1 x \sin{\phi} \label{d1-2_ellip}, \\
\partial_1\alpha_{1\Sigma} & = & a_{\Sigma}b_{\Sigma}  \left(J_0 + 2 K_0 x^2 \cos^2{\phi}\right), \label{d11_ellip} \\
\partial_2\alpha_{2\Sigma} & = & a_{\Sigma}b_{\Sigma}  \left(J_1 + 2 K_2 x^2 \cos^2{\phi}\right),\label{d22_ellip} \\
\partial_1\alpha_{2\Sigma} & = & a_{\Sigma}b_{\Sigma}  K_1 x \sin{2\phi}, \label{d12_ellip}
\end{eqnarray}
where $a_\Sigma$ and $b_\Sigma$ are defined in Eq. (\ref{coord_enfw}), 
\begin{equation}
J_n := \int_0^1 \frac{\kappa(m(u))}{[1 - (1 - b^2_\Sigma)u]^{1/2 + n}[1 - (1 - a^2_\Sigma)u]^{3/2 - n}}
\end{equation}
and 
\begin{equation}
K_n := \int_0^1 \frac{u\kappa^\prime(m(u))}{[1 - (1 - b^2_\Sigma)u]^{1/2 + n}[1 - (1 - a^2_\Sigma)u]^{5/2 - n}},
\end{equation}
where $\kappa$ is the convergence of the circular model, $\kappa^\prime=\frac{1}{2m}\frac{d\kappa(m)}{dm}$ and we have defined the variable $m^2=x^2g(u,\phi)$  such that
\begin{displaymath}
g^2(u,\phi) = u \left(   \frac{\cos^2\phi}{1 - (1-a_{\Sigma}^2)u} + \frac{\sin^2\phi}{1 - (1-b_{\Sigma}^2)u} \right).
\end{displaymath}

For the ENFW model, with the approximation (\ref{kappa_nfw}), we can rewrite the potential derivatives as
\begin{equation}
J_n(x,\phi)=\mathcal{J}_n\kappa(x)-2\kappa^\Sigma_s\mathcal{L}_n(\phi), \quad K_n(x,\phi)= -\frac{\kappa^\Sigma_s}{x}\mathcal{K}_n(\phi), \label{JK}
\end{equation}
where
\begin{eqnarray}
\mathcal{J}_n &=& \int_0^1 \frac{du}{[1 - (1 - b^2_\Sigma)u]^{1/2 + n}[1 - (1 - a^2_\Sigma)u]^{3/2 - n}}, \label{an_enfw}\\
\mathcal{K}_n(\phi) &=&\int_0^1 \frac{ug(\phi,u)^{-2}du}{[1 - (1 - b^2_\Sigma)u]^{1/2 + n}[1 - (1 - a^2_\Sigma)u]^{5/2 - n}}. \label{dn_enfw} \\
\mathcal{L}_n(\phi) &=&\int_0^1 \frac{\log{g(\phi, u) du }}{[1 - (1 - b^2_\Sigma)u]^{1/2 + n}[1 - (1 - a^2_\Sigma)u]^{3/2 - n}}, \label{bn_enfw} \label{Ln}
\end{eqnarray}

Substituting the expressions above in Eq. (\ref{d1-2_ellip}) we obtain Eqs. (\ref{alpha1_enfw}) and (\ref{alpha2_enfw}). Similarly, by substituting Eqs. (\ref{JK}) -- (\ref{Ln})  in  (\ref{d11_ellip}) -- (\ref{d12_ellip}) and
using the definitions  (\ref{conv}) and (\ref{shear1_2}) we obtain Eqs. (\ref{kappa_enfw}) -- (\ref{g2_enfw}).

\section{Fitting functions \label{app_fit}} 
Applying the procedure outlined in Sect. \ref{lim_cdc}, we obtained the maximum values of $\varepsilon$ ($\varepsilon_\Sigma$) as a function of $\kappa^\varphi_s$ ($\kappa^\Sigma_s$) such that the analytic solutions derived in the paper are accurate. We find that these functions are well fitted by a Pad\'e approximant of the form

\begin{equation}
\tilde{\varepsilon}(\kappa_s)=\frac{\sum_{n=0}^{4}a_n(\kappa_s)^n}{\sum_{m=0}^{2}b_m(\kappa_s)^m},
\label{emax_fit_funct}
\end{equation}
where $\tilde{\varepsilon}$ and $\kappa_s$ correspond either to $\varepsilon,\kappa^\varphi_s$ (for the PNFW) or $\varepsilon_\Sigma,\kappa^\Sigma_s$ (for the ENFW), and the values of coefficients are given in Table \ref{fit_param}

\begin{table}[!ht]
\centering
  \begin{tabular}{|c|c|c|}
\hline
  &  PNFW   &  ENFW    \\
\hline
  $a_0$& $  0.187 $   &   $  0.300 $ \\
  $a_1$& $ -0.770 $   &   $ -1.236 $ \\
  $a_2$& $ -0.556 $   &   $ -1.481 $ \\
  $a_3$& $ -3.129 $   &   $ -0.796 $ \\
  \hline
  $b_0$& $ -0.024 $   &   $  0.125 $ \\
  $b_1$& $  1.670 $   &   $  0.542 $ \\
  $b_2$& $  5.021 $   &   $  4.717 $ \\
\hline
$\chi^2$ & $4.64\times 10^{-6}$& $8.69\times 10^{-6}$ \\
\hline
\end{tabular}
\caption{\label{fit_param} Results from the regression analysis using the Pad\'e approximant for $\tilde{\varepsilon}(\kappa_s)$. The last row corresponds to the values of $\chi^2$ for each function.}
\end{table}

\end{appendix}


\end{document}